\documentclass[lettersize,journal]{IEEEtran}
\usepackage{textcomp}
\usepackage{stfloats}
\usepackage{url}
\usepackage{verbatim}
\usepackage[T1]{fontenc}
\usepackage{ifpdf}

\usepackage{cite}

\usepackage[dvips]{graphicx}
\graphicspath{{../}}
\usepackage{graphicx}
\DeclareGraphicsExtensions{.pdf,.jpeg,.png,.eps}

\usepackage{amsmath}
\usepackage{amsfonts}
\usepackage{amssymb}
\usepackage{mathrsfs}
\usepackage[cmintegrals]{newtxmath}
\usepackage{bm}

\newtheorem{example}{\textbf{Example}}

\usepackage[ruled,linesnumbered]{algorithm2e}
\usepackage{array}
\usepackage{multirow}
\usepackage[table,xcdraw]{xcolor}
\usepackage{booktabs,threeparttable}
\usepackage[caption=false,font=normalsize,labelfont=sf,textfont=sf]{subfig}

\begin{document}
	\title{Low-Complexity PSCL Decoding \\of Polar Codes}
	\author{Xinyuanmeng~Yao,~\IEEEmembership{Member,~IEEE} and Xiao~Ma,~\IEEEmembership{Member,~IEEE}
	\thanks{Xinyuanmeng Yao is with the School of Cyber Science and Engineering, Ningbo University of Technology, Ningbo 315211, China, and was with the School of Computer Science and Engineering and Guangdong Province Key Laboratory of Information Security Technology, Sun Yat-sen University, Guangzhou 510006, China (e-mail:yaoxym@nbut.edu.cn) .
		
	Xiao Ma is with the School of Computer Science and Engineering, and also with Guangdong Province Key Laboratory of Information Security Technology, Sun Yat-sen University, Guangzhou 510006, China (e-mail: maxiao@mail.sysu.edu.cn). The corresponding author is Xiao Ma.}
	}
	\maketitle
	\begin{abstract}
	Successive cancellation list~(SCL) decoding enables polar codes and their generalizations to deliver satisfactory performance in finite-length scenarios but it comes with high latency and complexity. To reduce latency, a partitioned SCL~(PSCL) decoding algorithm, implemented over a PSCL decoding tree, can be utilized. In this work, we aim to lower down the complexity of the PSCL decoding, resulting in an efficient decoding algorithm with low latency and complexity for polar-like codes. To achieve this, we define two metrics at each level of the PSCL decoding tree. One is for evaluating the reliability of a path and the other is for evaluating the reliability of a list of paths. Then, we propose a double-threshold strategy in the PSCL decoding process where unreliable valid paths are pruned based on the first metric, and then the smallest reliable list of paths is selected to continue based on the second metric. Simulation results demonstrate that when polar/CRC-polar/PAC codes are decoded using the proposed low-complexity PSCL decoder, both the sorting complexity and the computational complexity are reduced and significantly decrease as the signal-to-noise ratio (SNR) increases.
	\end{abstract}
	
	
	\section{Introduction}\label{section1}
	Polar codes are a family of error-correcting codes which can provably achieve the capacities of binary-input discrete memoryless channels~(BI-DMCs) in an asymptotic sense~\cite{Arikan2009}. However, practical polar codes of short to moderate lengths exhibit inefficiency under successive cancellation (SC) decoding. Notably, the SC performance of finite-length polar codes is inferior to the performance achieved by other well-established codes such as Bose-Chaudhuri-Hocquenghem~(BCH) codes, low-density parity-check~(LDPC) codes and turbo codes~\cite{Coskun2019}. To enhance the error-correcting capabilities of polar codes, Tal and Vardy proposed SC list~(SCL) decoding in~\cite{Tal2011,Tal2015}. Subsequent research also shows that under SCL decoding, the polar code generalizations, such as cyclic redundancy check (CRC)-polar codes~\cite{Niu2012} and polarization-adjusted convolutional~(PAC) codes~\cite{Yao2021}, can achieve superior performance compared to the original polar codes.
	
	It is widely accepted that there are two main concerns associated with SCL. Firstly, SCL requires a bit-by-bit advancement in the decoding process, leading to high latency. In order to address this issue, researchers proposed to design multi-bits parallel decoders for special nodes. In works such as~\cite{Alamdar-Yazdi2011,Sarkis2014,Hanif2017,Hashemi2016,Hashemi2017,Hanif2018,Condo2018,Ardakani2019}, special nodes are identified based on the positions of active and frozen bits. 
	{ A more general approach to identify special nodes is based on the number of active bits~\cite{Yao2024}, which allows us to use the so-called partitioned SCL (PSCL) decoding\footnote{The paradigm of PSCL decoding was proposed in~\cite{Hashemi2016ICASSP}, where a polar code is partitioned into constituent codes that are decoded separately. In~\cite{Hashemi2016ICASSP}, the partitioning is based on a length constraint, resulting in constituent codes of equal length. In contrast, \cite{Yao2024} employs a dimension-based partitioning approach, producing constituent codes with typically different lengths.}~\cite{Hashemi2016ICASSP}.}
	Second, during the SCL decoding process, each decoding path will be extended into two paths when decoding an information bit. If the number of the extended paths exceeds a preset list size of $L$, the SCL decoder will sort all the paths and select the most probable $L$ paths to continue, which leads to high complexity. To mitigate the complexity, it is necessary to design a strategy to efficiently discard redundant candidate paths at each decoding step without sacrificing performance. According to the timing of the discard operation, we can classify existing related works into two categories: discard before sorting, e.g.~\cite{Zhang2016,Gao2019,Wang2021,Moradi2023}, and discard after sorting, e.g.~\cite{Chen2013,Chen2016}. 
	
	Considering the aforementioned drawbacks associated with the conventional SCL decoder, we design a low-complexity PSCL decoder in this work, which is implemented over a PSCL decoding tree. On the one hand, we introduce a novel reliability metric for an extended path. At each decoding step, all the invalid paths, along with some unreliable paths, are first pruned. On the other hand, we introduce a metric for a given list of extended paths, which is to evaluate the probability of the correct path being included in the list at each decoding step. After the pruning step, we select a small but reliable list of paths to continue.
	
	The remainder of this paper is organized as follows. In Section~\ref{section2}, we first give a detailed description of the SC/SCL decoding based on the polar tree and the SCL decoding tree, followed by an introduction to the PSC/PSCL decoding based on the sub-polar tree and the PSCL decoding tree. In Section~\ref{section3}, we define two crucial metrics and provide the related simulation analysis. Subsequently, we introduce a novel low-complexity PSCL decoding algorithm. Simulation results are shown in Section~\ref{section4}. Finally, Section~\ref{section5} offers a brief summary of the paper.
	
	\section{Preliminaries}\label{section2}
	\subsection{Notations}
	In this paper, we use uppercase letters, e.g., $X$, and lowercase letters, e.g., $x$, to denote random variables and their realizations, respectively. An $\ell$-length random vector and its realization are expressed as $\boldsymbol{X}=(X_1,X_2,\ldots,X_{\ell})$ and $\boldsymbol{x}=(x_1,x_2,\ldots,x_{\ell})$, respectively. For simplicity, we use $[N]$ to denote the set $\{1,2,\ldots,N\}$. For a subset $\mathcal{A}\subseteq[N]$, we use $\mathcal{A}^c$ to represent the complement of $\mathcal{A}$, i.e., $\mathcal{A}^c=[N]\backslash \mathcal{A}$. Given an $N$-length vector $\boldsymbol{x}$, $\boldsymbol{x}_{\mathcal{A}}$ represents the subvector of $\boldsymbol{x}$, $(x_{i_1},x_{i_2},\ldots,x_{i_{|\mathcal{A}|}})$, where $i_1<i_2<\ldots<i_{|\mathcal{A}|}$ enumerate the elements in $\mathcal{A}$ whose cardinality is $|\mathcal{A}|$. A matrix is expressed as $\mathbf{X}$. In addition, the binary field is denoted by $\mathbb{F}_2$ and the real field is denoted by $\mathbb{R}$.
	
	\subsection{Polar Tree}
\begin{figure}[!t]
		\centering
		\includegraphics[width=0.48\textwidth]{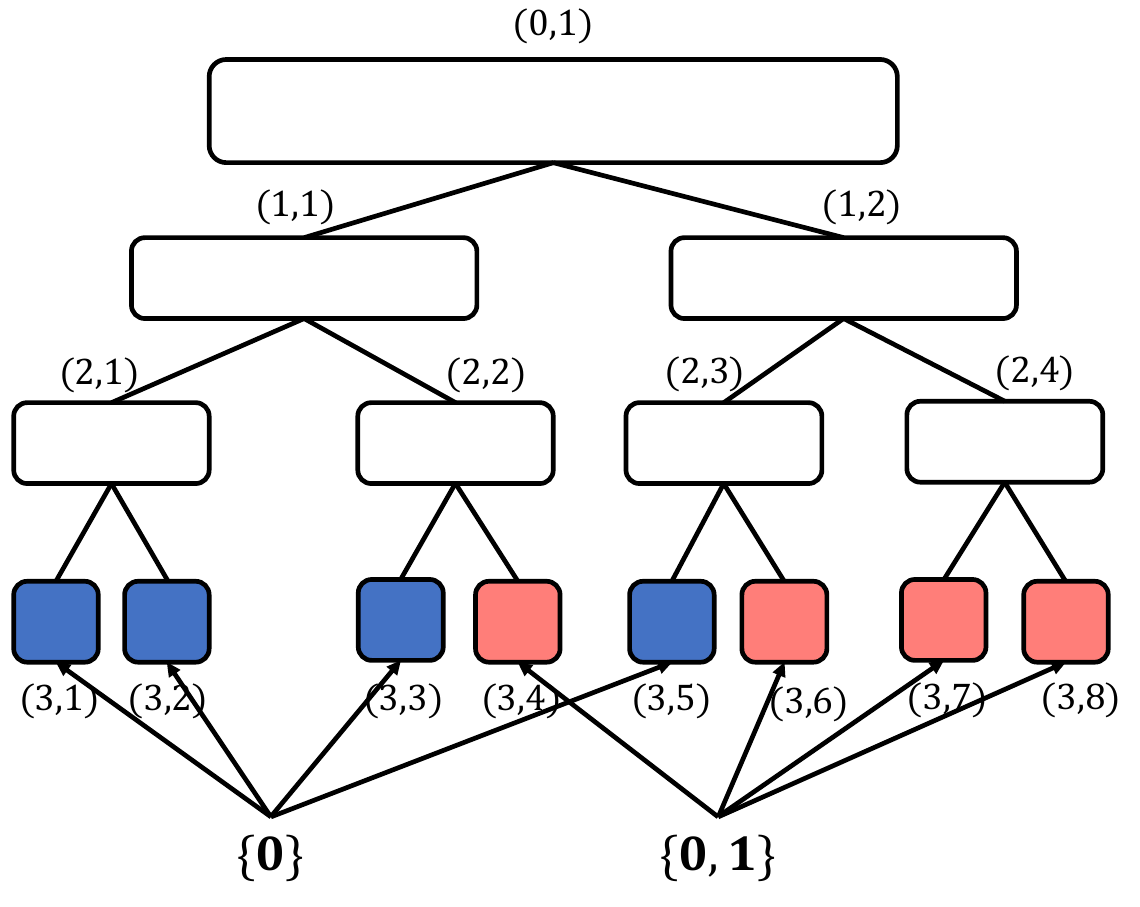}
		\caption{The polar tree for $\text{Polar}(8,4,\{4,6,7,8\})$. The blue leaf nodes correspond to the set $\{0\}$ and the red leaf nodes correspond to the set $\{0,1\}$.}\label{PolarTree}
	\end{figure}

	Given three parameters, block length $N=2^n$, dimension $K$ and information index set $\mathcal{A}\subseteq[N]$ with $|\mathcal{A}|=K$, we can construct a polar tree with $n+1$ levels. The root node is at the $0$-th level. For $0\leq s<n$, each node at the $s$-th level has two children at the $(s+1)$-th level, resulting in $N$ leaf nodes at the $n$-th level. Furthermore, each node is indexed by a two-tuple $(s,t)$, where $s$~($0\leq s\leq n$) represents the level and $t$~($1\leq t\leq 2^s$) represents the position of the node from left to right at the $s$-th level, and has a length $\ell^{(s,t)}$ as well as a set of bit sequences with length $\ell^{(s,t)}$, denoted by  $\mathcal{C}^{(s,t)}$. Here, the lengths are given by $\ell^{(s,t)}=2^{n-s}$ and the sets $\mathcal{C}^{(s,t)}$ are defined as follows. For a leaf node $(n,t)$, its corresponding set is $\mathcal{C}^{(n,t)}=\{0,1\}$ if $t\in\mathcal{A}$ and $\mathcal{C}^{(n,t)}=\{0\}$ if $t\in\mathcal{A}^c$. For a non-leaf node $(s,t)$ ($s<n$), its corresponding set is  $\mathcal{C}^{(s,t)}=\{(\boldsymbol{a}\oplus \boldsymbol{b}, \boldsymbol{b})|\boldsymbol{a}\in\mathcal{C}^{(s+1,2t-1)},\boldsymbol{b}\in\mathcal{C}^{(s+1,2t)}\}$, where $\oplus$ denotes the vector addition over $\mathbb{F}_2$. Meanwhile, we obtain a polar code, which is exactly the set $\mathcal{C}^{(0,1)}$, and denote it by $\text{Polar}(N,K,\mathcal{A})$. In Fig.~\ref{PolarTree}, we draw the polar tree for $\text{Polar}(8,4,\{4,6,7,8\})$ for illustration.
	
	\subsection{Polar Coding}
	At the transceiver, the encoding and decoding of $\text{Polar}(N,K,\mathcal{A})$ can be implemented over the corresponding polar tree. 
	\subsubsection{Encoding}
	Assume that each node on the polar tree has a sequence, and denote by $\boldsymbol{u}^{(s,t)}=(u_1^{(s,t)},u_2^{(s,t)},\ldots,u_{\ell(s,t)}^{(s,t)})$ the $\ell(s,t)$-length sequence of the node $(s,t)$. For encoding, we first map $K$ information bits to $u_1^{(n,t)}$, $t\in\mathcal{A}$, and set $u_1^{(n,t)}$, $t\in\mathcal{A}^c$ to zeros, and then calculate the sequences of all the inner nodes recursively from the leaves to the root according to $\boldsymbol{u}^{(s,t)}=(\boldsymbol{u}^{(s+1,2t-1)}\oplus \boldsymbol{u}^{(s+1,2t)}, \boldsymbol{u}^{(s+1,2t)})$. Let $\boldsymbol{v}\triangleq(u_1^{(n,1)},u_1^{(n,2)},\ldots,u_1^{(n,N)})$ be the data-carrier sequence and $\boldsymbol{c}\triangleq\boldsymbol{u}^{(0,1)}$ be the transmitted codeword. The encoding process can be algebraically expressed as $\boldsymbol{c}=\boldsymbol{v}\mathbf{G}^{\otimes n}$, where $\mathbf{G}=(\begin{smallmatrix}1~0\\1~1\end{smallmatrix})$ and $\otimes$ denotes the Kronecker product.
	
	Throughout this paper, we consider the binary-input additive white Gaussian noise channel~(BI-AWGNC). The polar codeword $\boldsymbol{c}$ is mapped to a channel input $\boldsymbol{x}$ as $x_i = (-1)^{c_i},c_i\in\mathbb{F}_2$, and then transmitted over an AWGNC, resulting in a channel output $\boldsymbol{y}\in\mathbb{R}^N$. The BI-AWGNC has the channel transition probability density function given by
	\begin{align}
		W(y_i|x_i)=\frac{1}{\sqrt{2\pi\sigma^2}}\exp[-(y_i-x_i)^2/(2\sigma^2)],
	\end{align}
	where $\sigma^2$ is the variance of the zero-mean Gaussian noise. The signal-to-noise ratio~(SNR) is defined as $E_b/N_0$, where $E_b$ is the energy per data bit and $N_0=2\sigma^2$.
	
	\subsubsection{Successive Cancellation Decoding}
	Assume that each node on the polar tree has a log-likelihood ratio~(LLR) sequence and a hard bit estimate~(HBE) sequence. Denote by $\boldsymbol{\alpha}^{(s,t)}=(\alpha_1^{(s,t)},\alpha_2^{(s,t)},\ldots,\alpha_{\ell(s,t)}^{(s,t)})$ and $\boldsymbol{\beta}^{(s,t)}=(\beta_1^{(s,t)},\beta_2^{(s,t)},\ldots,\beta_{\ell(s,t)}^{(s,t)})$ the $\ell(s,t)$-length LLR sequence and the $\ell(s,t)$-length hard bit estimate sequence of the node $(s,t)$, respectively. Define 
	\begin{equation}\label{Function:f}
		f(a,b)\triangleq\ln\frac{1+e^{a+b}}{e^a+e^b},
	\end{equation} 
	for $a,b\in\mathbb{R}$, and
	\begin{equation}\label{Function:g}
		g(a,b,c)\triangleq b+(-1)^c\cdot a,
	\end{equation}
	for $a,b\in\mathbb{R}$ and $c\in\mathbb{F}_2$ .
	Given a channel output $\boldsymbol{y}$, we first calculate the LLR sequence of the root node, and then proceed according to the following four rules.  
	\begin{enumerate}
		\item When the LLR sequence of a parent node $\boldsymbol{\alpha}^{(s,t)}$ is available, the LLR sequence of its left child $\boldsymbol{\alpha}^{(s+1,2t-1)}$ is calculated as
		\begin{align}
			\alpha_i^{(s+1,2t-1)}=f(\alpha^{(s,t)}_i,\alpha^{(s,t)}_{i+2^{n-s-1}}),
		\end{align}
		for $i=1,2,\ldots,2^{n-s-1}$.
		\item When the LLR sequence of a parent node $\boldsymbol{\alpha}^{(s,t)}$ and the HBE sequence of its left child $\boldsymbol{\beta}^{(s+1,2t-1)}$ are available, the LLR sequence of its right child $\boldsymbol{\alpha}^{(s+1,2t)}$ is calculated as
		\begin{equation}
			\alpha^{(s+1,2t)}_i=g(\alpha^{(s,t)}_i,\alpha^{(s,t)}_{i+2^{n-s-1}},\beta^{(s+1,2t-1)}_i),
		\end{equation}
		for $i=1,2,\ldots,2^{n-s-1}$.
		\item For a leaf node $(n,t)$, when its LLR $\alpha_1^{(n,t)}$ is available, its HBE is calculated as
		\begin{align}
			\beta_1^{(n,t)}=
			\begin{cases}
				1,~t\in\mathcal{A} \text{~and~} \alpha_1^{(n,t)}<0 \\
				0,~\text{otherwise}
			\end{cases}.
		\end{align}
		\item When the HBE sequences of a pair of sibling nodes,  $\boldsymbol{\beta}^{(s+1,2t-1)}$ and $\boldsymbol{\beta}^{(s+1,2t)}$, are available, the HBE sequence of their parent $\boldsymbol{\beta}^{(s,t)}$ is calculated as
		\begin{equation}
			\boldsymbol{\beta}^{(s,t)}=(\boldsymbol{\beta}^{(s+1,2t-1)}\oplus\boldsymbol{\beta}^{(s+1,2t)},\boldsymbol{\beta}^{(s+1,2t)}).
		\end{equation}
	\end{enumerate}
	Once all the LLR sequences and the HBE sequences are obtained, the decoding process is terminated and the estimate of the data-carrier sequence $\hat{\boldsymbol{v}}$ satisfies $\hat{\boldsymbol{v}}=\boldsymbol{\beta}^{(0,1)}\mathbf{G}^{\otimes n}$. 
	
	\subsubsection{Successive Cancellation List Decoding}
	The SCL decoding can enhance the error performance of polar codes, whose process can be graphically represented by a path search on a tree. We name the tree as the SCL decoding tree, as defined below. 
	
\begin{figure}[!t]
		\centering
		\includegraphics[width=0.43\textwidth]{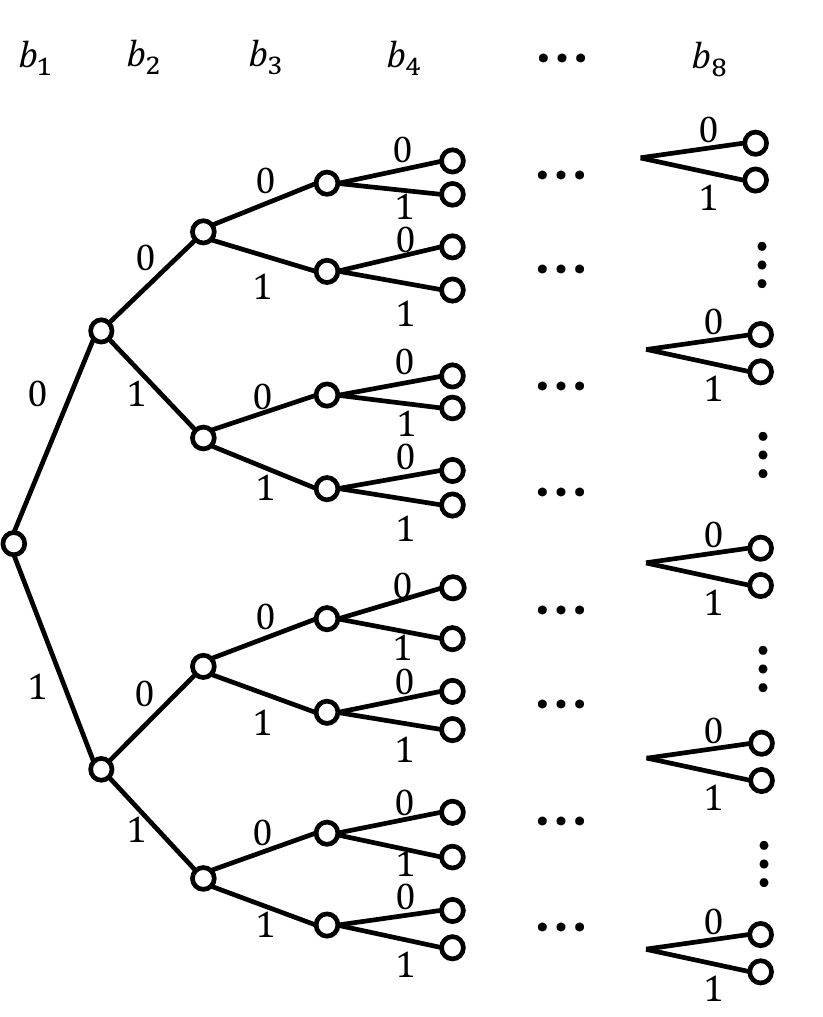}
		\caption{The SCL decoding tree for $\text{Polar}(8,4,\{4,6,7,8\})$. }\label{SCLDecodingTree}
	\end{figure}

	For $\text{Polar}(N,K,\mathcal{A})$, the corresponding SCL decoding tree has $N+1$ levels. The root node is at the $0$-th level. For $0\leq r<N$, each node at the $r$-th level has two children at the $(r+1)$-th level, resulting in $2^N$ leaf nodes at the $N$-th level. By convention, we label the upper branch and the lower branch by 0 and 1, respectively. Then, for each path which starts from the root node and ends at a node at the $r$-th level, we can use a $r$-bit sequence $(b_1,\ldots,b_r)$ to represent it and assign a metric as
	\begin{align}
		Q^{(r)}(b_1,\ldots,b_r)\triangleq
		\sum\limits_{1\leq i \leq r}\ln(1+e^{-(1-2b_i)\cdot\alpha_1^{(n,i)}}),
	\end{align}
	where $\alpha_1^{(n,i)}$ is the LLR of the $i$-th leaf node during the SC decoding when the HBEs of the first $i-1$ leaf nodes $\beta_1^{(n,1)},\ldots,\beta_1^{(n,i-1)}$ are set to $b_1,\ldots,b_{i-1}$, respectively. Note that for the root node, it is represented by $\emptyset$ and has  $Q^{(0)}(\emptyset)=0$.  As an example, the SCL decoding tree for $\text{Polar}(8,4,\{4,6,7,8\})$ is shown in Fig.~\ref{SCLDecodingTree}.
	
	Then, upon receiving a channel output $\boldsymbol{y}$, we commence at the root node and extend paths from the $0$-th level to the $N$-th level over the SCL decoding tree. In practice, we keep at most $L$ paths at each level. To this end, we proceed according to the following two rules.
	\begin{enumerate}
		\item At the $r$-th level, where $r\in\mathcal{A}^c$, prune the invalid paths $(b_1,\ldots,b_r)$ with $b_r=1$.
		\item At the $r$-th level, where $r\in\mathcal{A}$, if there are more than $L$ paths remained, sort all the remaining paths and select the $L$ paths with the minimum metrics $Q^{(r)}(b_1,\ldots,b_r)$ to survive.
	\end{enumerate}
	
	Eventually, the path $(b_1,\ldots,b_N)$ with the minimum metric $Q^{(N)}(b_1,\ldots,b_N)$ is selected as the SCL decoding output, i.e., $\hat{\boldsymbol{v}}=(b_1,\ldots,b_N)$. 
	
	\subsection{Partitioned Successive Cancellation-Based Decoding}
	The SC-based decoding requires the decoding process to advance bit by bit, which results in high latency. To reduce the decoding latency, the PSC-based decoding was proposed, which can be implemented over a sub-polar tree. 
	\subsubsection{Sub-Polar Tree}
\begin{figure}[!t]
		\centering
		\includegraphics[width=0.45\textwidth]{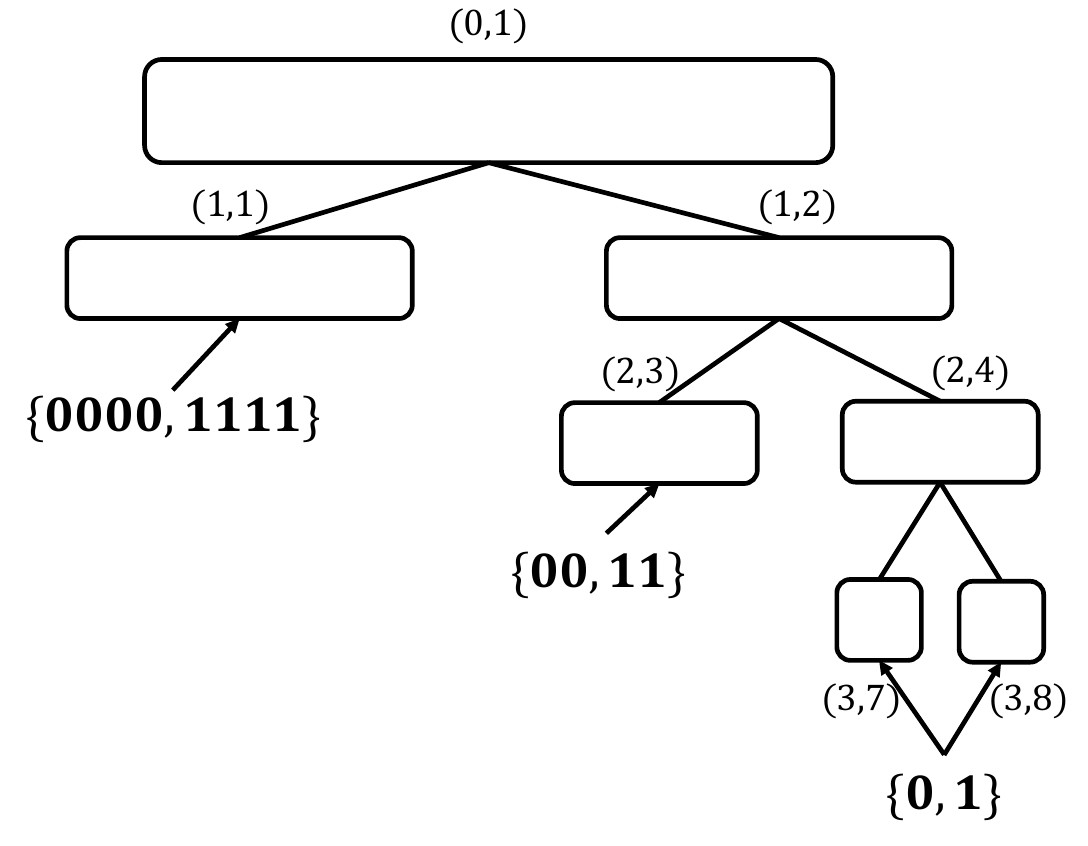}
		\caption{The sub-polar tree with $\tau=1$ for $\text{Polar}(8,4,\{4,6,7,8\})$.}\label{SubPolarTree}
	\end{figure}

	Let $\tau$~(a positive integer) be a dimension threshold. For a tree node, if its corresponding set has cardinality no greater than $2^{\tau}$ and the corresponding set of its parent has cardinality greater than $2^{\tau}$, then we remove all the descendants of the node. In this way, we can extract a sub-polar tree from the original polar tree. In Fig.~\ref{SubPolarTree}, the sub-polar tree with $\tau=1$ for $\text{Polar}(8,4,\{4,6,7,8\})$ is depicted.
	
	Suppose that there are $M$ leaf nodes on the sub-polar tree. For ease of notation, we relabel the $M$ leaf nodes using $\mathcal{M}[i]$ to represent the $i$-th leaf node from left to right, where $i$ ranges from $1$ to $M$ and $\mathcal{M}[i]$ represents some $(s,t)$. Note that on the sub-polar tree, most of the leaves are not at the $n$-th level and have length greater than one.
	\subsubsection{Partitioned Successive Cancellation Decoding}
	For PSC decoding, we also assume that each node on the sub-polar tree has a LLR sequence and a HBE sequence. The PSC decoding procedure is similar to the SC procedure, except for the third rule. In the PSC decoding, for a leaf node $\mathcal{M}[r]$ at the sub-polar tree, when its LLR $\boldsymbol{\alpha}^{\mathcal{M}[r]}$ is available, its HBE is calculated as
	\begin{align}
		\boldsymbol{\beta}^{\mathcal{M}[r]}=
		\mathop{\arg \max}\limits_{(a_1,a_2,\ldots,a_{\ell(\mathcal{M}[r])})\in\mathcal{C}^{\mathcal{M}[r]}} \sum_{i=1}^{\ell(\mathcal{M}[r])}(1-2a_i)\alpha_i^{\mathcal{M}[r]}.
	\end{align}
	Obviously, the PSC decoding output $\hat{\boldsymbol{v}}$ can be obtained according to $(\boldsymbol{\beta}^{\mathcal{M}[1]},\boldsymbol{\beta}^{\mathcal{M}[2]},\ldots,\boldsymbol{\beta}^{\mathcal{M}[M]})$.
	
	\subsubsection{Partitioned Successive Cancellation List Decoding}
\begin{figure}[!t]
		\centering
		\includegraphics[width=0.28\textwidth]{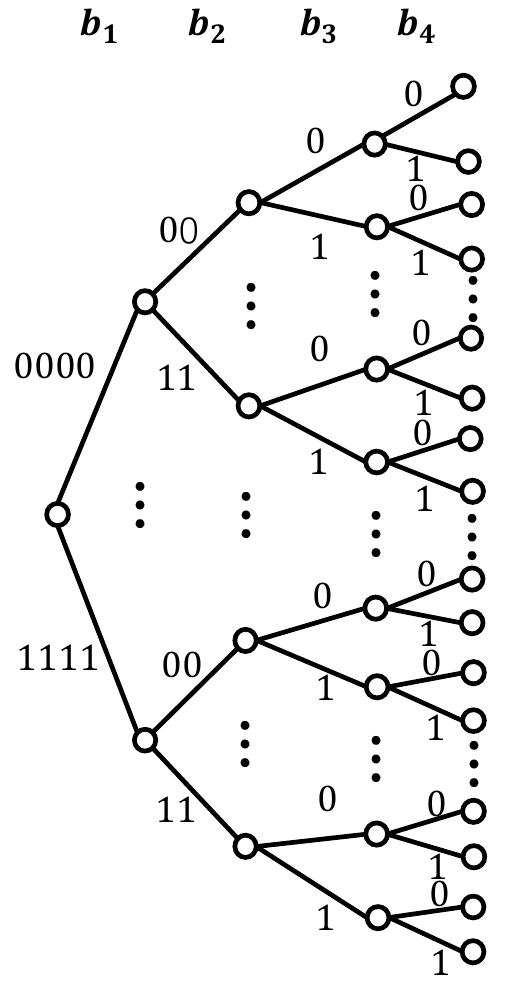}
		\caption{The PSCL decoding tree with $\tau=1$ for $\text{Polar}(8,4,\{4,6,7,8\})$. }\label{PSCLDecodingTree}
	\end{figure}

	To depict the PSCL decoding, we construct a PSCL decoding tree. The PSCL decoding tree has $M+1$ levels. The root node is at the $0$-th level. For $0\leq r<M$, each node at the $r$-th level has $2^{\ell(\mathcal{M}[r+1])}$ children at the $(r+1)$-th level, resulting in $2^N$ leaf nodes at the $M$-th level. Progressing from top to bottom, the $i$-th branch emitting from a node at the $r$-th level is labeled by the $\ell(\mathcal{M}[r+1])$-length binary representation of $(i-1)$, where $i=1,2,\ldots,2^{\ell(\mathcal{M}[r+1])}$. Then, for each path which starts from the root node and ends at a node at the $r$-th level~($0\leq r \leq M$), we can use a $r$-segment sequence $(\boldsymbol{b}_1,\ldots,\boldsymbol{b}_r)$ to represent it and assign a metric as
	\begin{align}
		Q^{(r)}(\boldsymbol{b}_1,\ldots,\boldsymbol{b}_r)
		\triangleq
		\sum\limits_{1\leq i \leq r}\sum\limits_{1\leq j \leq \ell(\mathcal{M}[i])}\ln(1+e^{-(1-2b_{i,j})\cdot\alpha_j^{\mathcal{M}[i]}}),
	\end{align}
	where $b_{i,j}$ denotes the $j$-th component of $\boldsymbol{b}_i$ and $\alpha_j^{\mathcal{M}[i]}$ is the $j$-th LLR of the $i$-th leaf node during the PSC decoding when the HBEs of the first $r$ leaf nodes $\boldsymbol{\beta}^{\mathcal{M}[1]},\ldots,\boldsymbol{\beta}^{\mathcal{M}[r]}$ are set to $\boldsymbol{b}_1,\ldots,\boldsymbol{b}_r$, respectively. As an example, the PSCL decoding tree with $\tau=1$ for $\text{Polar}(8,4,\{4,6,7,8\})$ is shown in Fig.~\ref{PSCLDecodingTree}. 
	
	Then, upon receiving a channel output $\boldsymbol{y}$, we commence at the root node and extend paths from the $0$-th level to the $M$-th level over the PSCL decoding tree. At the $r$-th level~($1\leq r\leq M$), we perform two steps:
	\begin{enumerate}
		\item \textit{pruning step:} prune invalid paths $(\boldsymbol{b}_1,\ldots,\boldsymbol{b}_r)$ with $\boldsymbol{b}_r\notin\mathcal{C}^{\mathcal{M}[r]}$;
		\item \textit{selection step:} if there are more than $L$ paths remained, sort all the remaining paths and select the $L$ paths with the minimum metrics $Q^{(r)}(\boldsymbol{b}_1,\ldots,\boldsymbol{b}_r)$ to survive.
	\end{enumerate}
	
	Eventually, the PSCL decoding output $\hat{\boldsymbol{v}}$ can be obtained according to the path $(\boldsymbol{b}_1,\ldots,\boldsymbol{b}_M)$ with the minimum metric $Q^{(M)}(\boldsymbol{b}_1,\ldots,\boldsymbol{b}_M)$. 
	
	\section{A Low-Complexity PSCL Decoding Algorithm}\label{section3}
	As demonstrated in~\cite{Yao2024}, the PSCL decoding can match the performance of the SCL decoding while offering reduced latency. However, the sorting complexity of PSCL decoding significantly increases with larger values of the dimension threshold $\tau$ and/or the list size $L$.
	
	Given a PSCL decoding tree, let $l_r$ denote the number of the surviving paths at the $r$-th level and $e_r$ denote the number of the valid branches from a node at the $r$-th level. To guarantee that at most $L$ reliable paths are survived at the $(r+1)$-th level, the PSCL decoder needs to sort $l_r\times e_r$ paths. Based on this observation, we introduce a double-threshold strategy in this section, which aims to reduce $e_r$ by pruning more paths before the sorting operation and $l_r$ by selecting less paths to survive after the sorting operation. 
	
	\subsection{The Pruning Step}
	In the pruning step, we propose to prune not only the invalid paths but also the unreliable valid paths. To this end, we first define a metric to evaluate the reliability of a path.
	
	At the $r$-th level~($1\leq r\leq M$), define a metric for a path $(\boldsymbol{b}_1,\ldots,\boldsymbol{b}_r)$ as
	\begin{align}
		R^{(r)}(\boldsymbol{b}_1,\ldots,\boldsymbol{b}_r)
		\triangleq
		\frac{1}{\ell(\mathcal{M}[r])}\sum_{j=1}^{\ell(\mathcal{M}[r])}(-1)^{b_r,j}\alpha_j^{\mathcal{M}[r]},
	\end{align}
	where $\alpha_j^{\mathcal{M}[r]}$ is the $j$-th LLR of the $r$-th leaf node during the PSC decoding when the HBEs of the first $(r-1)$ leaf nodes $\boldsymbol{\beta}^{\mathcal{M}[1]},\ldots,\boldsymbol{\beta}^{\mathcal{M}[r-1]}$ are set to $\boldsymbol{b}_1,\ldots,\boldsymbol{b}_{r-1}$, respectively. 
	\begin{example}
		Consider the example with $N=128,K=64,\tau=2$ and set $E_b/N_0=2~\text{dB}$. For $r\in\{1,3,4\}$, the histograms of $R^{(r)}(\boldsymbol{b}_1,\ldots,\boldsymbol{b}_r)$ with the correct  $(\boldsymbol{b}_1,\ldots,\boldsymbol{b}_{r-1})$ are shown in Fig.~\ref{GAAnalysisN128K64Tau2_AvgLLR}. We observed that $R^{(r)}(\boldsymbol{b}_1,\ldots,\boldsymbol{b}_r)$ with $\boldsymbol{b}_r$ being the correct one~(marked in red) is likely to be large. 
		\begin{figure*}[!t]
			\centering
			\includegraphics[width=\textwidth]{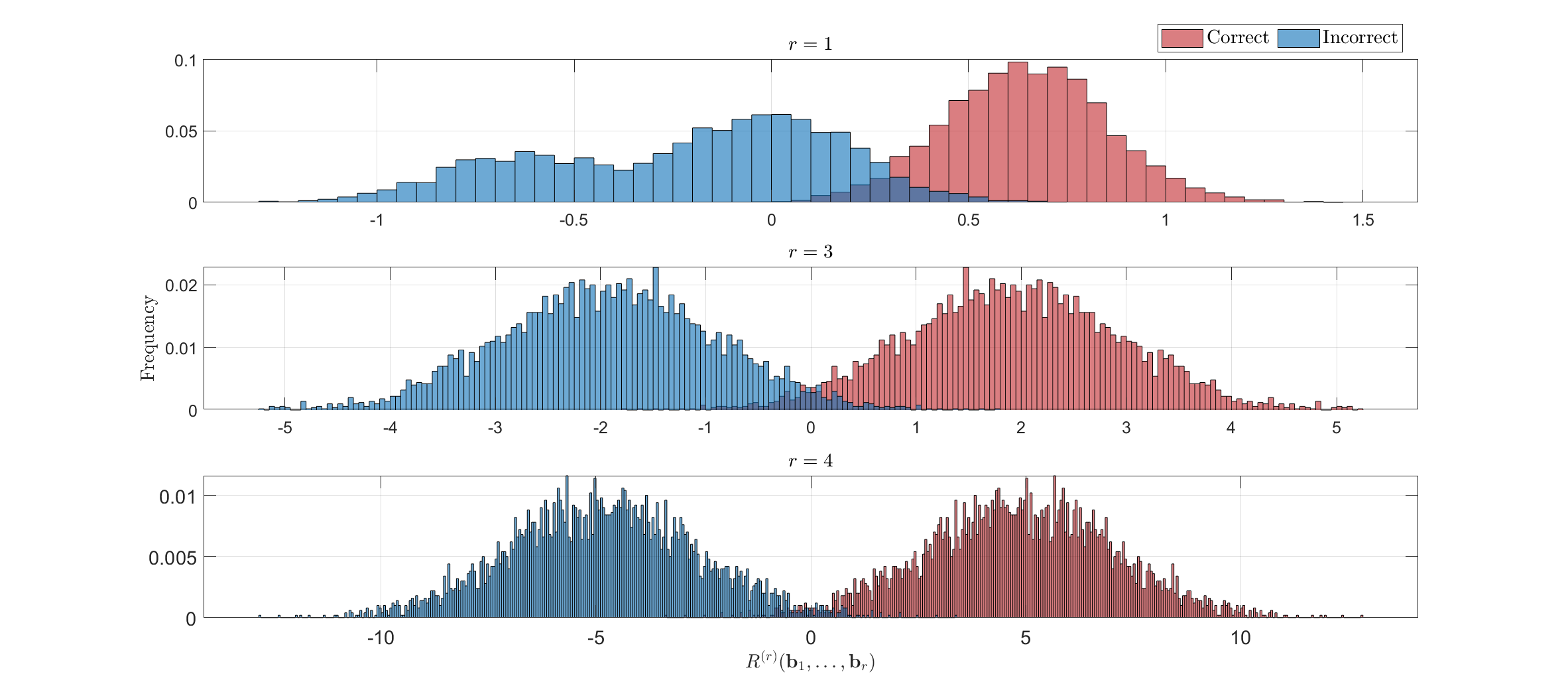}
			\caption{Statistical behavior of $R^{(r)}(\boldsymbol{b}_1,\ldots,\boldsymbol{b}_r)$. }\label{GAAnalysisN128K64Tau2_AvgLLR}
		\end{figure*}
	\end{example}
	
	The statistical behavior of  $R^{(r)}(\boldsymbol{b}_1,\ldots,\boldsymbol{b}_r)$ motivates us to set a pruning threshold for the $r$-level, denoted by $\eta_{\text{pruning}}^{(r)}$. Then, besides the invalid paths, the valid paths with $R^{(r)}(\boldsymbol{b}_1,\ldots,\boldsymbol{b}_r)<\eta_{\text{pruning}}^{(r)}$ can be considered as unreliable and thus pruned as well.
	
	To determine $\eta_{\text{pruning}}^{(r)}$, we propose the following method. Consider that the all-zero codeword is transmitted over a BI-AWGNC and assume that the LLRs of the tree nodes at the sub-polar tree are Gaussian random variables with mean $\mu$ and variance $2\mu$. For a given SNR, we can use the density evolution method to obtain the distribution of the random variable associated with $\alpha_j^{\mathcal{M}[r]}$ when the HBEs of the first $(r-1)$ leaf nodes are set correctly, and denote it by ${\rm{\mathcal{N}}}(\mu^{(r)},2\mu^{(r)})$. Then, it can be deduced that for the correct path $(\boldsymbol{b}_1,\ldots,\boldsymbol{b}_r)$, its corresponding $R^{(r)}(\boldsymbol{b}_1,\ldots,\boldsymbol{b}_r)$ follows a distribution given by $R^{(r)}\sim{\rm{\mathcal{N}}}(\mu^{(r)},2\mu^{(r)}/\ell(\mathcal{M}[r]))$. To illustrate the accuracy of the approximation, we provide an example below. Therefore, we can select  $\eta_{\text{pruning}}^{(r)}$ such that $\Pr\{R^{(r)}<\eta_{\text{pruning}}^{(r)}\}\leq \varepsilon_{tol}$, where $\varepsilon_{tol}$ denotes a tolerable error probability.
	
	\begin{example}
	Consider the same case in the first example. For $r\in\{1,3,4\}$, the cumulative distribution functions~(CDFs) of $R^{(r)}$ obtained by the Monte-Carlo simulation and the GA-based method are compared in Fig.~\ref{GAAnalysisN128K64Tau2_AvgLLRDistribution}, from which we see that the CDF curves from the GA-based method closely approximate those from the Monte-Carlo simulation.
		\begin{figure*}[!t]
			\centering
			\includegraphics[width=0.8\textwidth]{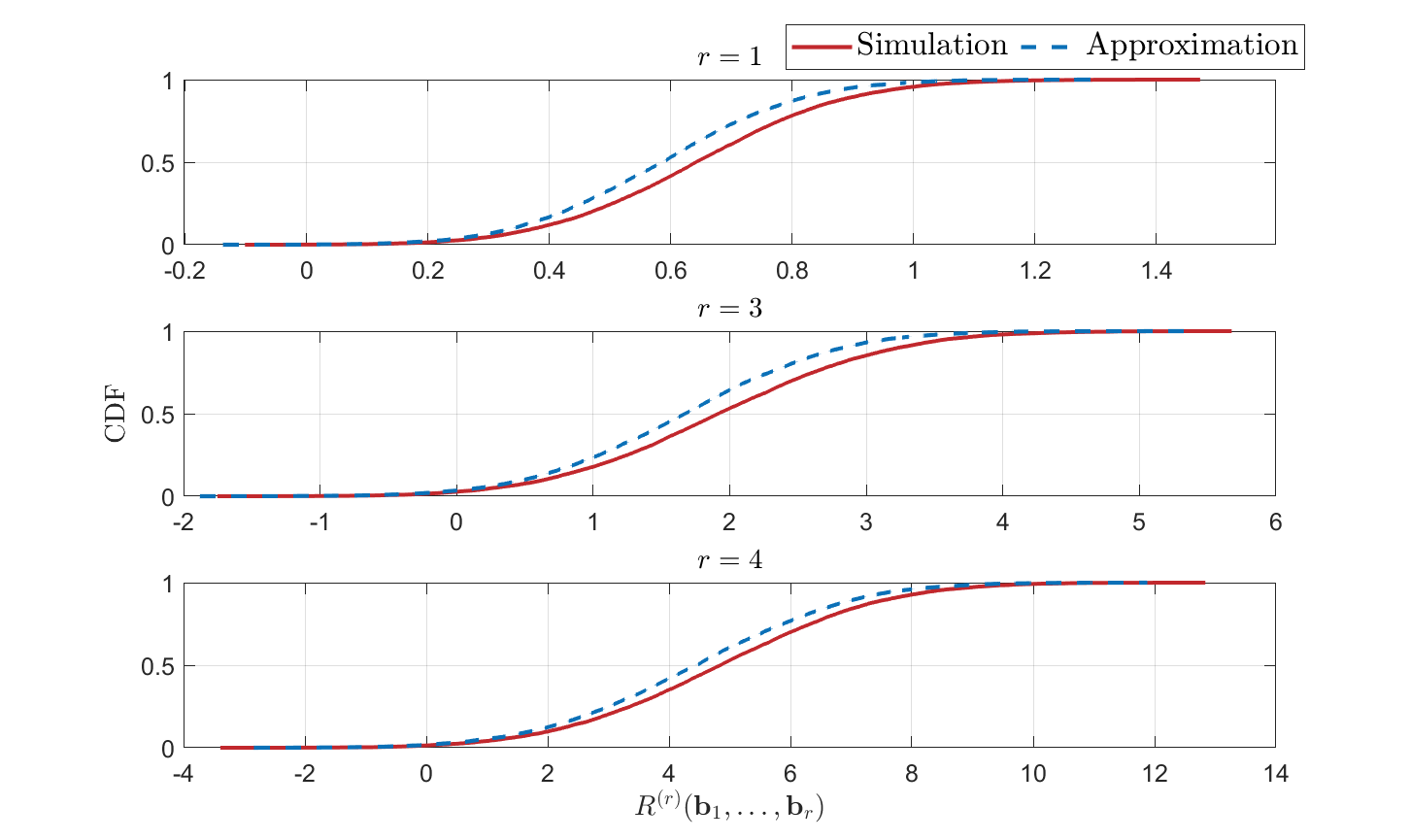}
			\caption{The CDFs of $R^{(r)}(\boldsymbol{b}_1,\ldots,\boldsymbol{b}_r)$ obtained by the Monte-Carlo simulation and the GA-based approximation method.}\label{GAAnalysisN128K64Tau2_AvgLLRDistribution}
		\end{figure*}
	\end{example}
	
	\subsection{The Selection Step}
   {
    In the selection step, we propose to select as few paths as possible without discarding the correct path. To this end, we define a metric to evaluate the reliability of a list of paths~(i.e., the probability that the correct path is in a list of paths).
    
    Consider the PSCL decoding procedure. Let $\widetilde{Q}^{(r)}\left(b_1, \ldots, b_r\right)$ represent the myopic probability~(likelihood) that the encoding sequences of the first $r$ leaf nodes on the sub-polar tree are $\boldsymbol{b}_1, \ldots, \boldsymbol{b}_r$, which is defined as
    \begin{align}
    	\widetilde{Q}^{(r)}\left(\boldsymbol{b}_1, \ldots, \boldsymbol{b}_r\right)=e^{Q^{(r)}\left(\boldsymbol{b}_1, \ldots, \boldsymbol{b}_r\right)}.
    \end{align}
    Then, we have
    \begin{align}
    	\widetilde{Q}^{(r)}(\boldsymbol{b}_1,\ldots,\boldsymbol{b}_r)=\sum_{(\boldsymbol{a}_1,\ldots,\boldsymbol{a}_M):\atop\boldsymbol{a}_i=\boldsymbol{b}_i,1\leq i \leq r }\widetilde{Q}^{(M)}(\boldsymbol{a}_1,\ldots,\boldsymbol{a}_M).
    \end{align}
    Notice that the above summation is over all~(valid or invalid) $M$-length complete paths. Let $\mathcal{U}$ denote the set of all valid $M$-length complete paths. Then, for a valid $r$-length path $(\boldsymbol{b}_1,\ldots,\boldsymbol{b}_r)$, we define a function
    \begin{align}
    	\phi(\boldsymbol{b}_1,\ldots,\boldsymbol{b}_r)\triangleq\sum_{(\boldsymbol{a}_1,\ldots,\boldsymbol{a}_M)\in\mathcal{U}:\atop\boldsymbol{a}_i=\boldsymbol{b}_i,1\leq i \leq r }\widetilde{Q}^{(M)}(\boldsymbol{a}_1,\ldots,\boldsymbol{a}_M),
    \end{align}
    which represents the sum of likelihoods for all valid $M$-length complete paths $(\boldsymbol{a}_1, \ldots, \boldsymbol{a}_M)$ originating from $(\boldsymbol{b}_1, \ldots, \boldsymbol{b}_r)$. Further, without abuse of notation, for a set of valid $r$-length paths $\mathcal{U}^{(r)}$, we can define
    \begin{align}
    	\phi(\mathcal{U}^{(r)})\triangleq\sum_{(\boldsymbol{b}_1,\ldots,\boldsymbol{b}_r)\in\mathcal{U}^{(r)}}\phi(\boldsymbol{b}_1,\ldots,\boldsymbol{b}_r).
    \end{align}
    
    Suppose that at the $j$-th decoding step of the PSCL decoding, the set of valid $j$-length paths $\mathcal{W}^{(j)}$ is discarded, and at the $r$-th decoding step, the set of valid $r$-length paths $\mathcal{V}^{(r)}$ is retained, where $1\leq j\leq r$. To evaluate the reliability of $\mathcal{V}^{(r)}$, we introduce the following metric:
    \begin{align}
    	\Gamma^{(r)}(\mathcal{V}^{(r)})\triangleq\frac{\phi(\mathcal{V}^{(r)})}{\phi(\mathcal{V}^{(r)})+
    		\sum_{j=1}^{r}\phi(\mathcal{W}^{(j)})}.
    \end{align}
    With the above notations, we see that $\Gamma^{(r)}(\mathcal{V}^{(r)})$ is exactly the a posteriori probability that the correct path is included in $\mathcal{V}^{(r)}$. Intuitively, the greater the metric $\Gamma^{(r)}(\mathcal{V}^{(r)})$ is, the higher probability that the correct path is retained. Unfortunately, the exact computations of $\phi(\mathcal{V}^{(r)})$ and $\phi(\mathcal{W}^{(j)})~(1\leq j\leq r)$ are intractable at this time. Thus we turn to similar approximations adopted in~\cite{Yuan2024}. Specifically,  we assume that, for a given $\ell$-length path $(\boldsymbol{b}_1,\ldots,\boldsymbol{b}_{\ell})$, any two $M$-length complete paths $(\boldsymbol{a}_1,\ldots,\boldsymbol{a}_M)$ and $(\boldsymbol{a}'_1,\ldots,\boldsymbol{a}'_M)$ satisfying $a_i=a'_i=b_i$ for $1\leq i\leq \ell$ have equal myopic probabilities, i.e., $\widetilde{Q}^{(M)}(\boldsymbol{a}_1,\ldots,\boldsymbol{a}_M)=\widetilde{Q}^{(M)}(\boldsymbol{a}'_1,\ldots,\boldsymbol{a}'_M)$. Under this assumption, we can use $\widetilde{Q}^{(\ell)}(\boldsymbol{b}_1,\ldots,\boldsymbol{b}_\ell)$ to approximate $\phi(\boldsymbol{b}_1,\ldots,\boldsymbol{b}_{\ell})$ as
    \begin{align}
    	\phi(\boldsymbol{b}_1,\ldots,\boldsymbol{b}_{\ell})
    	\approx
    	\frac{\upsilon^{(\ell)}}{\omega^{(\ell)}}
    	\widetilde{Q}^{(\ell)}(\boldsymbol{b}_1,\ldots,\boldsymbol{b}_\ell),
    \end{align}
    where
    \begin{align}
    	\upsilon^{(\ell)}\triangleq|\{(\boldsymbol{a}_1,\ldots,\boldsymbol{a}_M)\in\mathcal{U}:\boldsymbol{a}_i=\boldsymbol{b}_i,1\leq i \leq \ell\}|,
    \end{align}
    and
    \begin{align}
    	\omega^{(\ell)}\triangleq|\{(\boldsymbol{a}_1,\ldots,\boldsymbol{a}_M):\boldsymbol{a}_i=\boldsymbol{b}_i,1\leq i \leq \ell\}|.
    \end{align}
    This leads to the approximations
    \begin{align}\label{Eq:ApproximationV}
    	\phi(\mathcal{V}^{(r)})\approx
    	\sum_{(\boldsymbol{b}_1,\ldots,\boldsymbol{b}_r)\in\mathcal{V}^{(r)}}\frac{\upsilon^{(r)}}{\omega^{(r)}}
    	\widetilde{Q}^{(r)}(\boldsymbol{b}_1,\ldots,\boldsymbol{b}_r),
    \end{align}	
    and
    \begin{align}\label{Eq:ApproximationW}
    	\phi(\mathcal{W}^{(j)})\approx\!\!\!\!\!\!\!\!
    	\sum_{(\boldsymbol{b}_1,\ldots,\boldsymbol{b}_j)\in\mathcal{W}^{(j)}}\frac{\upsilon^{(j)}}{\omega^{(j)}}
    	\widetilde{Q}^{(j)}(\boldsymbol{b}_1,\ldots,\boldsymbol{b}_j)\text{~for~}1\leq j\leq r.
    \end{align}
	
	Here, we provide a brief illustration of the approximation computation of  $\Gamma^{(r)}(\mathcal{V}^{(r)})$ in Example~\ref{Example:APSCLDecodingProcedure}. Additionally, to demonstrate the efficiency of $\Gamma^{(r)}(\mathcal{V}^{(r)})$, we present an example in Example~\ref{Example:LER}, employing an approach inspired by the work of~\cite{Yuan2024}. 
	\begin{figure}[!h]
		\centering
		\includegraphics[width=0.2\textwidth]{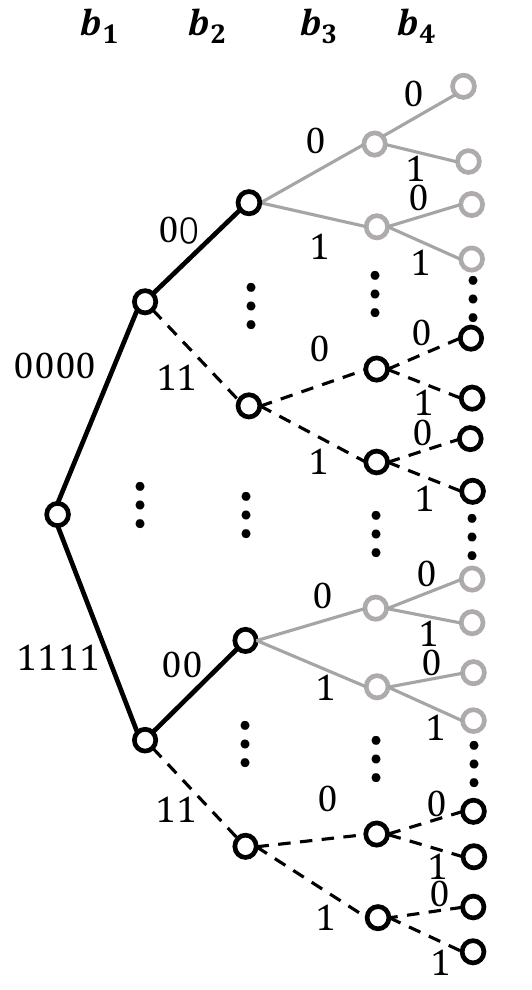}
		\caption{A PSCL decoding procedure with $\tau=1$ for $\text{Polar}(8,4,\{4,6,7,8\})$, where $L=2$. }\label{APSCLDecodingProcedure}
	\end{figure}
	\begin{figure}
		\centering
		\includegraphics[width=0.5\textwidth]{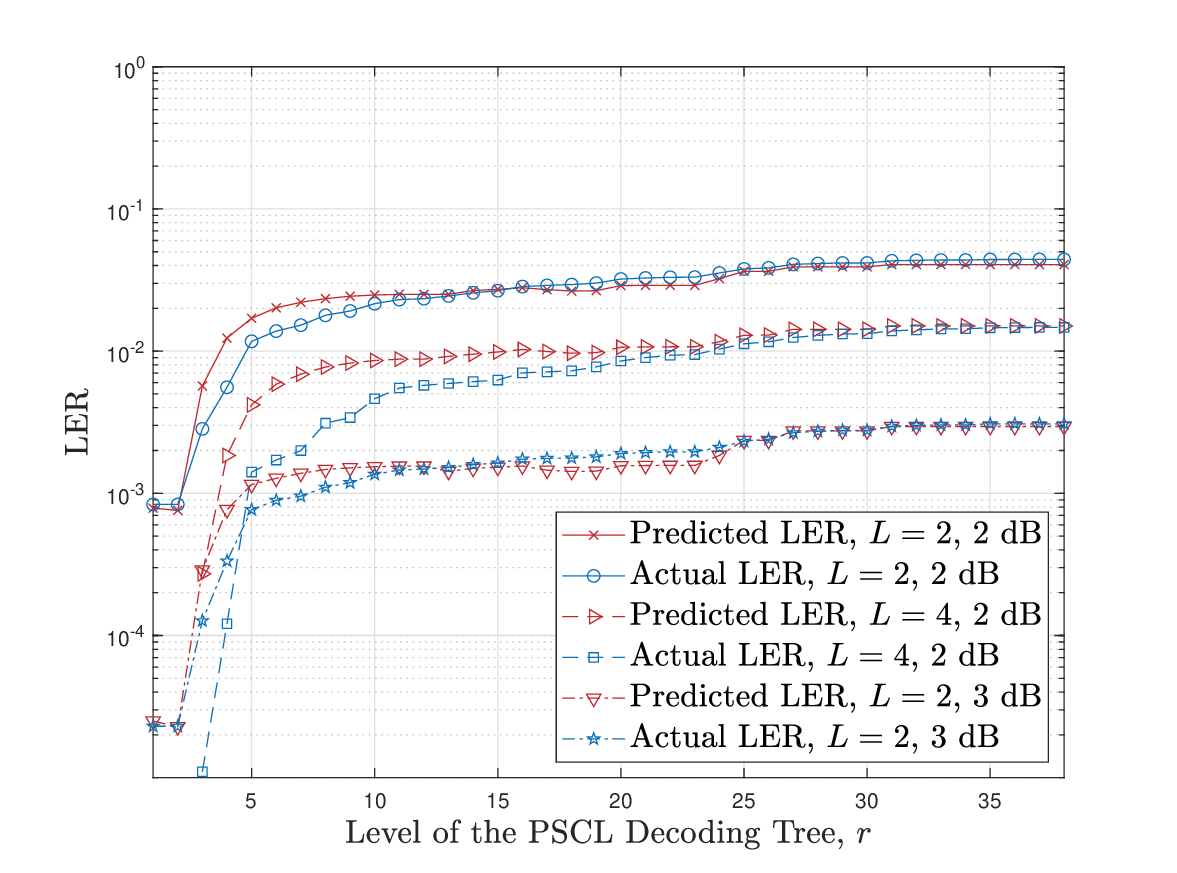}
		\caption{The actual LERs and the predicted LERs calculated by $\mathbb{E}[1-\Gamma^{(r)}(\mathcal{V}^{(r)})]$ under the PSCL decoding. Here, $N=128,K=64,\tau=2$.}\label{ListAnalysisN128K64Tau2}
	\end{figure}
	\begin{example}\label{Example:APSCLDecodingProcedure}
	In Fig.~\ref{APSCLDecodingProcedure}, we show a PSCL decoding procedure with $\tau=1$ for $\text{Polar}(8,4,\{4,6,7,8\})$, where $L=2$. At the first level, there are only two valid paths, $(0000,00)$ and $(1111,00)$~(marked with black solid lines), and both paths are preserved. Then, we have $\mathcal{V}^{(1)}=\{(0000),(1111)\}$ and $\mathcal{W}^{(1)}=\emptyset$. At the second level, there are four valid paths. Among them, two paths, $(0000,00)$ and $(1111,00)$, are preserved~(marked with black solid lines), while the other two paths, $(0000,11)$ and $(1111,11)$, are discarded~(marked with black dotted lines). Then, we have $\mathcal{V}^{(2)}=\{(0000,00),(1111,00)\}$ and $\mathcal{W}^{(2)}=\{(0000,11),(1111,11)\}$. At this time, to evaluate the reliability of $\mathcal{V}^{(2)}$, we compute $\Gamma(\mathcal{V}^{(2)})=\frac{\phi(\mathcal{V}^{(2)})}{\phi(\mathcal{V}^{(2)})+\phi(\mathcal{W}^{(1)})+\phi(\mathcal{W}^{(2)})}$. Since $\mathcal{W}^{(1)}=\emptyset$, we have $\phi(\mathcal{W}^{(1)})=0$. For $\phi(\mathcal{V}^{(2)})$, we use~(\ref{Eq:ApproximationV}) to approximate it as
	 \begin{align}
		\phi(\mathcal{V}^{(2)})\approx\sum_{\left(\boldsymbol{b}_1,\boldsymbol{b}_2\right) \in \mathcal{V}^{(2)}} \frac{4}{4} \widetilde{Q}^{(2)}\left(\boldsymbol{b}_1,\boldsymbol{b}_2\right),
	\end{align} 
	where the denominator ``4'' represents the number of the complete paths originating from $(\boldsymbol{b}_1,\boldsymbol{b}_2)\in\mathcal{V}^{(2)}$ and the numerator ``4'' represents the number of the valid complete paths originating from $(\boldsymbol{b}_1,\boldsymbol{b}_2)\in\mathcal{V}^{(2)}$. Similarly, we use~(\ref{Eq:ApproximationW}) to approximate $\phi(\mathcal{W}^{(2)})$ as:
	\begin{align}
		\phi(\mathcal{W}^{(2)})\approx\sum_{\left(\boldsymbol{b}_1,\boldsymbol{b}_2\right) \in \mathcal{W}^{(2)}} \frac{4}{4} \widetilde{Q}^{(2)}\left(\boldsymbol{b}_1,\boldsymbol{b}_2\right).
	\end{align}
	\end{example}
	\begin{example}\label{Example:LER}
	Consider the example with $N=128,K=64,\tau=2$. Define the list error rate~(LER) as the probability of the correct path not in the list. Fig.~\ref{ListAnalysisN128K64Tau2} plots the actual LERs as well as the predicted LERs calculated by $\mathbb{E}[1-\Gamma^{(r)}(\mathcal{V}^{(r)})]$ under the PSCL decoding. We can observe that the predicted LER is approximately monotonically increasing as $r$ increases and closes to the actual LER.
	\end{example}
	
	Based on the findings from Example~\ref{Example:LER}, we propose setting a global threshold of $\eta_{\text{selection}} \triangleq 1 - \varepsilon_{tol}$ and determine
	\begin{align}
		\mathcal{V}_{\text{min}}^{(r)}=\mathop{\arg\min}\limits_{\Gamma^{(r)}(\mathcal{V}^{(r)})\geq \eta_{\text{selection}}} |\mathcal{V}^{(r)}|
	\end{align}
	as the list of surviving paths in the selection step, where $\varepsilon_{tol}$ denotes a tolerable error probability.
}
	\subsection{The Proposed PSCL Decoding Algorithm}
	Now, we summarize the low-complexity PSCL decoding algorithm as follows.
	
	For a given SNR and a tolerable error probability $\varepsilon_{tol}$, we first calculate out $\eta_{\text{pruning}}^{(r)}(r=1,2,\ldots,M)$ and $\eta_{\text{selection}}$. Then, upon receiving a channel output $\boldsymbol{y}$, we commence at the root node and extend paths from the $0$-th level to the $M$-th level over the PSCL decoding tree. At the $r$-th level~($1\leq r\leq M$), we perform three steps:
	\begin{enumerate}
		\item \textit{Pruning Step:} Prune invalid paths with $\boldsymbol{b}_r\notin\mathcal{C}^{\mathcal{M}[r]}$ as well as valid but unreliable paths with $R_{\text{path}}^{(r)}(\boldsymbol{b}_1,\ldots,\boldsymbol{b}_r)<\eta_{\text{pruning}}^{(r)}$;
		\item \textit{Check Step:} Check if any paths remain after the pruning step. If not, terminate the decoding procedure prematurely.
		\item \textit{Selection Step:} Sort all the remaining paths and select a set of paths which is $\mathcal{V}_{\text{min}}^{(r)}=\mathop{\arg\min}\limits_{\Gamma^{(r)}(\mathcal{V}^{(r)})\geq \eta_{\text{selection}}} |\mathcal{V}^{(r)}|$. If there are more than $L$ paths in $\mathcal{V}_{\text{min}}^{(r)}$, select the $L$ paths with the minimum metrics $Q^{(r)}(\boldsymbol{b}_1,\ldots,\boldsymbol{b}_r)$ to survive.
	\end{enumerate}
	
	Eventually, the PSCL decoding output $\hat{\boldsymbol{v}}$ can be obtained according to the path $(\boldsymbol{b}_1,\ldots,\boldsymbol{b}_M)$ with the minimum metric $Q^{(M)}(\boldsymbol{b}_1,\ldots,\boldsymbol{b}_M)$. 
	
	\section{Simulation Results}\label{section4}
	In this section, simulation results are presented to illustrate the performance of the proposed low-complexity PSCL~(LC-PSCL) decoding algorithm. The error performance is measured by the frame error rate~(FER). The decoding complexity is analyzed from two perspectives: sorting complexity and computational complexity. The sorting complexity is measured by the average number of paths to be sorted  for decoding a codeword, while the computational complexity is measured by the average amount of floating-point operations~(FLOPs) required to decode a codeword, which are defined in~(\ref{Function:f}) or~(\ref{Function:g}). 
	
	For simplicity, we will use ``SCL($L$)'' to represent the conventional SCL decoding algorithm~\cite{Tal2015} with list size $L$, ``PSCL($L,\tau$)'' to represent the conventional PSCL decoding algorithm~\cite{Yao2024} with list size $L$ and dimension threshold $\tau$, and `` LC-PSCL($L,\tau$)'' to represent the proposed low-complexity PSCL decoding algorithm with list size $L$ and dimension threshold $\tau$. { The thresholds for LC-PSCL($L,\tau$), $\eta_{\text{pruning}}^{(r)}$ and $\eta_{\text{selection}}$, are determined using the following approach. For a given SNR, we first simulate the FER of PSCL($L,\tau$), denoted as $\varepsilon$, and then select a tolerable error probability formed as $\varepsilon_{tol} = \lambda \times \varepsilon$, ensuring that LC-PSCL($L,\tau$) maintains equivalent performance to PSCL($L,\tau$). Here, $\lambda=10^{-n}~(n\in\mathbb{N})$, is a tuning parameter used to adjust the tolerance level.}
	\begin{table}[!t]
	\caption{The parameters for the simulated polar codes}\label{TheParameters}
	\centering
	\begin{tabular}{@{}llll@{}}
		\toprule
		$N$ & $K$ & Rate Profile & $\lambda$ for LC-PSCL \\ \midrule
		128             & 32                 & 5G           & 0.001                        \\
		128             & 64                 & 5G           & 0.001                        \\
		128             & 96                 & 5G           & 0.001                        \\
		512             & 256                & 5G           & 0.0001                       \\
		128             & 64                 & Reed-Muller           & 0.001                        \\ \bottomrule
	\end{tabular}
    \end{table}
    \begin{figure}[!t]
	\begin{minipage}{\linewidth}
	\centering		\includegraphics[width=0.93\textwidth]{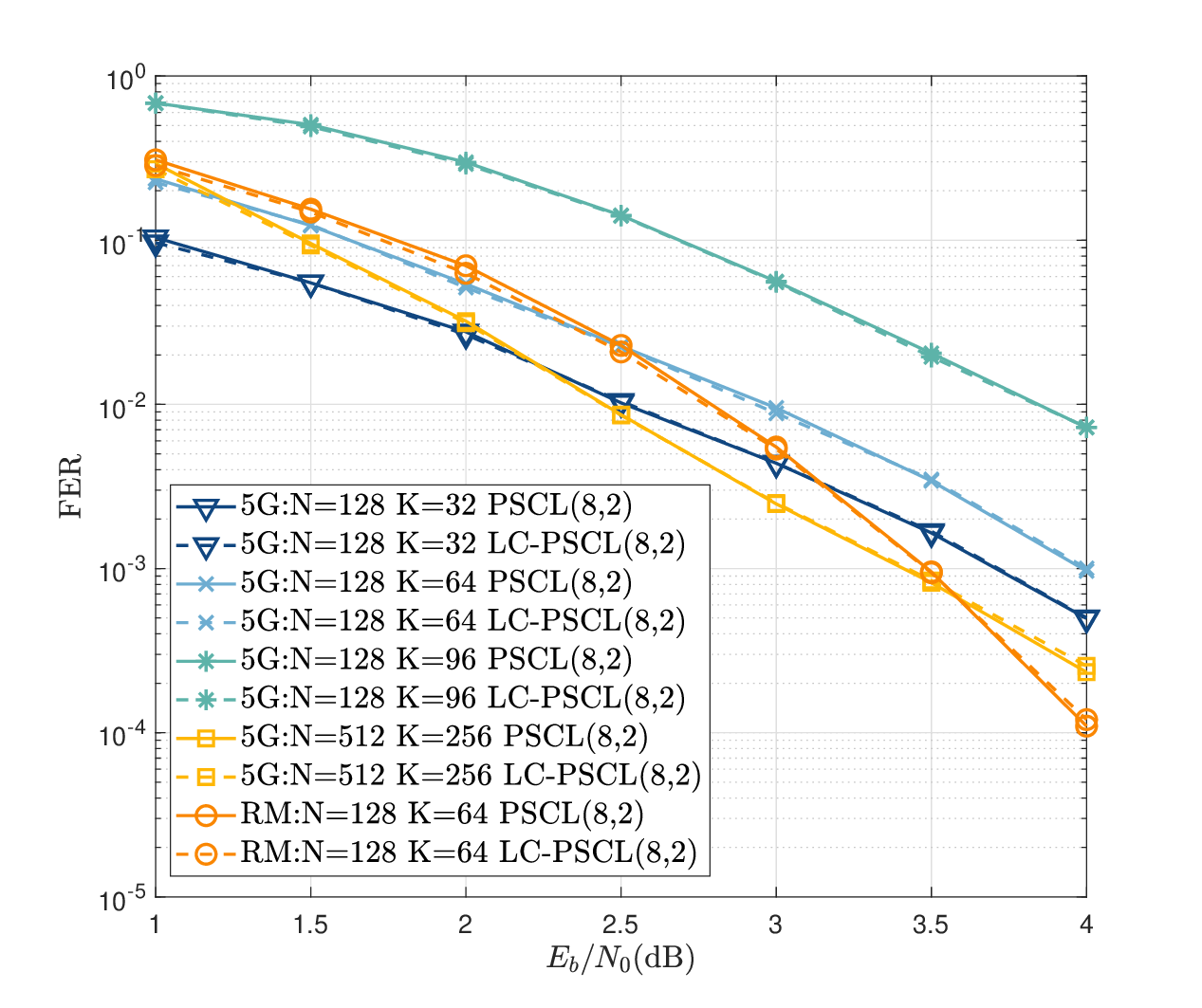}
	\caption{The error performance of different decoding algorithms for various polar codes. }\label{ebn0FER_DifferentParameters}	
	\end{minipage}
	
	\begin{minipage}{\linewidth}
	\centering		\includegraphics[width=0.93\textwidth]{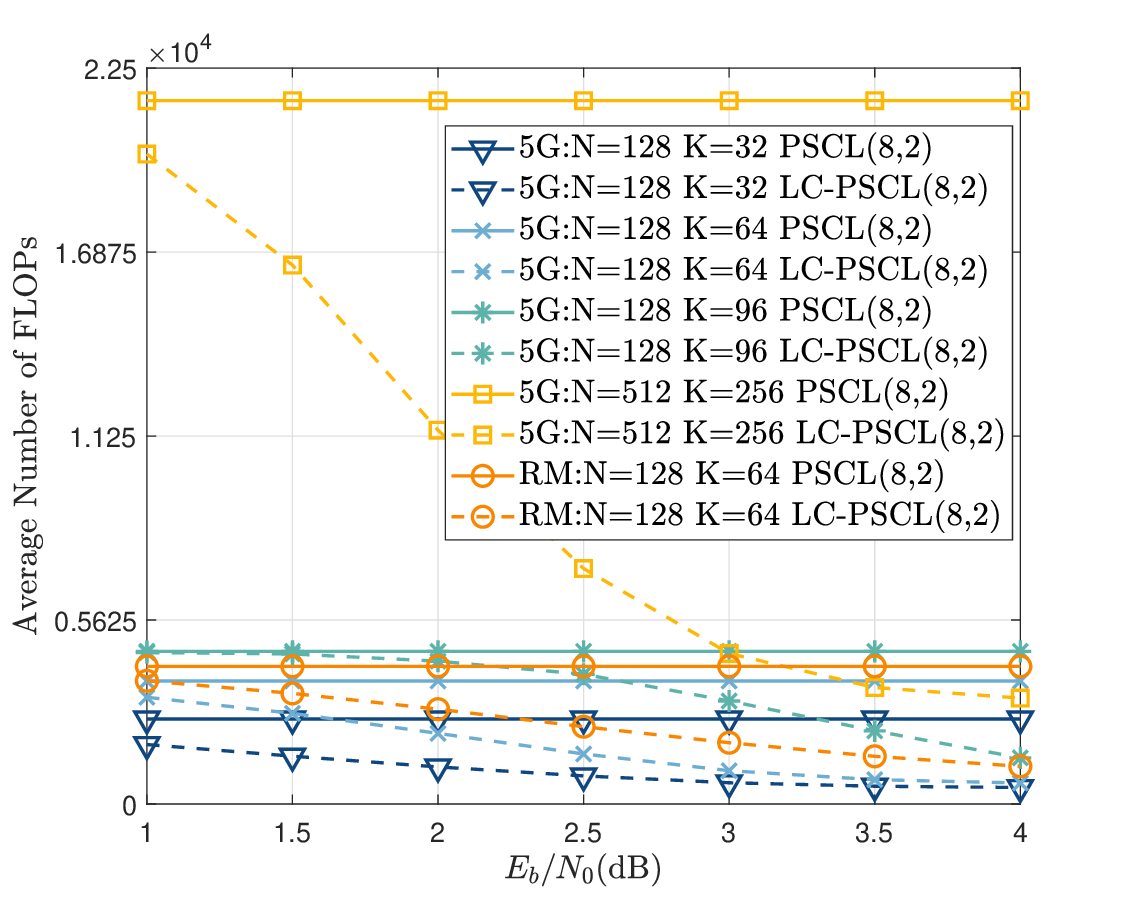}
	\caption{The computational complexity of different decoding algorithms for various polar codes. }\label{ebn0FLOP_DifferentParameters}
	\end{minipage}
	
	\begin{minipage}{\linewidth}
	\centering
	\includegraphics[width=0.93\textwidth]{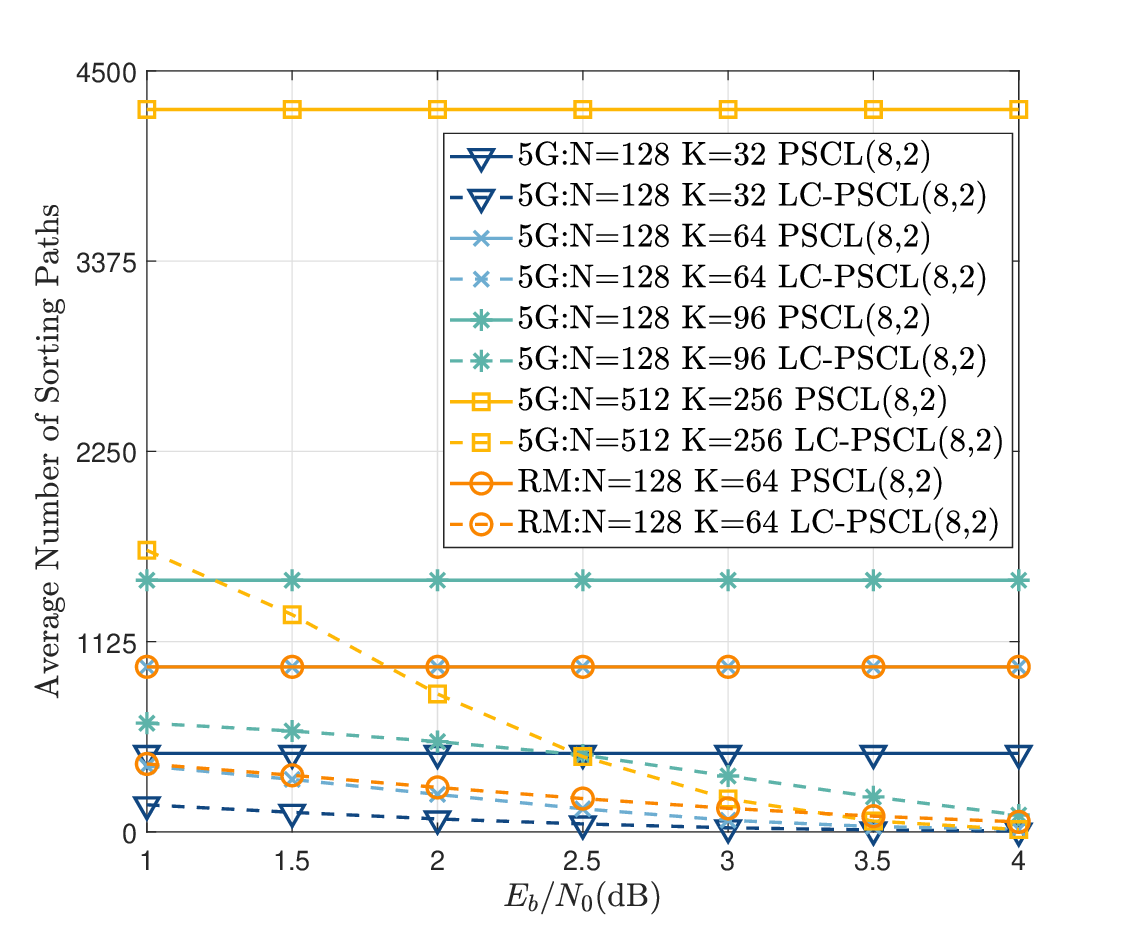}
	\caption{The sorting complexity of different decoding algorithms for various polar codes.}\label{ebn0SP_DifferentParameters}
	\end{minipage}
   \end{figure}		
	
	{
	\begin{example}
	To show the efficiency of our proposed LC-PSCL decoder, we simulate various polar codes with different code lengths, dimensions and rate profiles, and compare their PSCL decoding performance and LC-PSCL decoding performance. The parameters of the simulated polar codes are shown in Table~\ref{TheParameters}, and the corresponding simulation results are provided in Figs.~\ref{ebn0FER_DifferentParameters}-\ref{ebn0SP_DifferentParameters}. We observe that the tuning parameters $\lambda$ may vary for different polar codes. However, for any given polar code, provided the tuning parameter $\lambda$ is chosen appropriately, 
	\begin{enumerate}
	\item The LC-PSCL decoder can have no performance loss compared with the PSCL decoder, as shown in Fig.~\ref{ebn0FER_DifferentParameters}.
	\item The LC-PSCL decoder has significantly lower computational complexity and sorting complexity compared with the PSCL decoder, and the average number of FLOPs as well as the average number of sorting paths in the LC-PSCL decoding decreases as the SNR increases, as shown in Fig.~\ref{ebn0FLOP_DifferentParameters} and Fig.~\ref{ebn0SP_DifferentParameters}.
	\end{enumerate}
	\end{example}
     }
	\begin{figure}[!t]
	\begin{minipage}{\linewidth}	
	\centering
	\includegraphics[width=\textwidth]{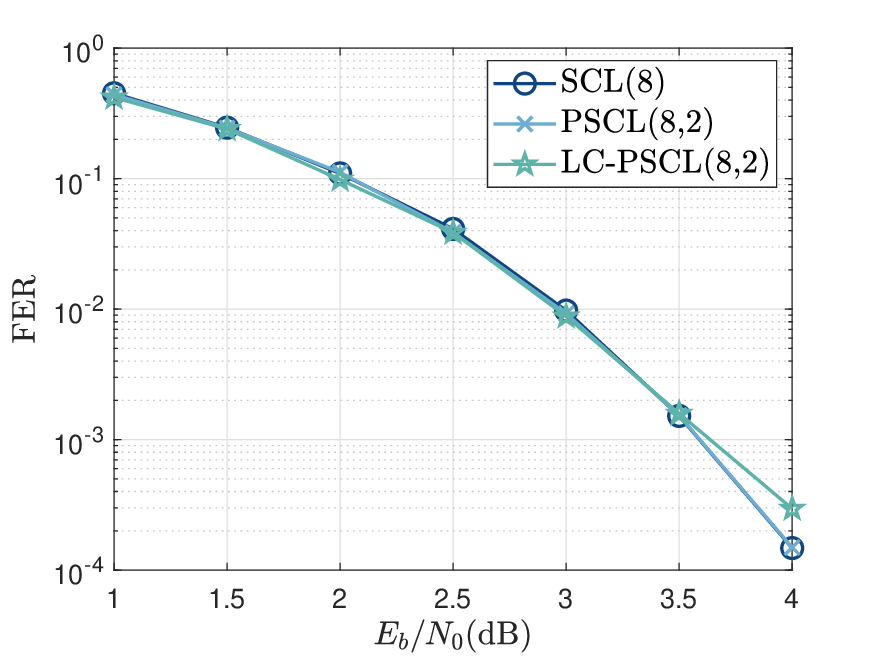}
	\caption{The error performance of different decoding algorithms for a CRC-polar code. Here, $N=128$ and $K=64$.}\label{ebn0FER_N128K64CRC11}
	\end{minipage}
	
	\begin{minipage}{\linewidth}
	\centering
	\includegraphics[width=\textwidth]{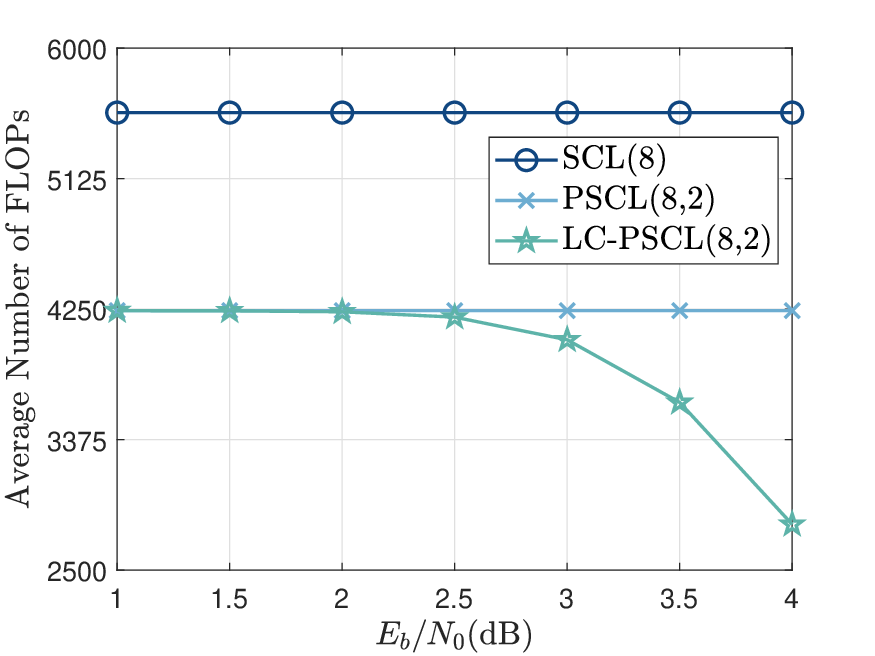}
	\caption{The computational complexity of different decoding algorithms for a CRC-polar code. Here, $N=128$ and $K=64$.}\label{ebn0FLOP_N128K64CRC11}
	\end{minipage}
	
	\begin{minipage}{\linewidth}
	\centering
	\includegraphics[width=\textwidth]{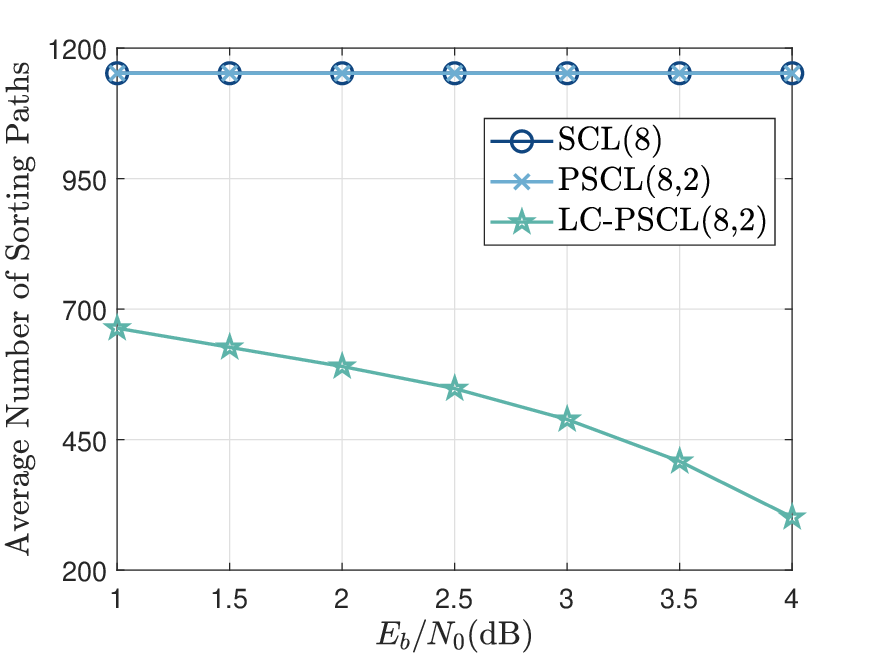}
	\caption{The sorting complexity of different decoding algorithms for a CRC-polar code. Here, $N=128$ and $K=64$.}\label{ebn0SP_N128K64CRC11}
	\end{minipage}
	\vskip0.1cm
\end{figure}	

    \begin{figure}[!t]
	\begin{minipage}{\linewidth}		
	\centering
	\includegraphics[width=\textwidth]{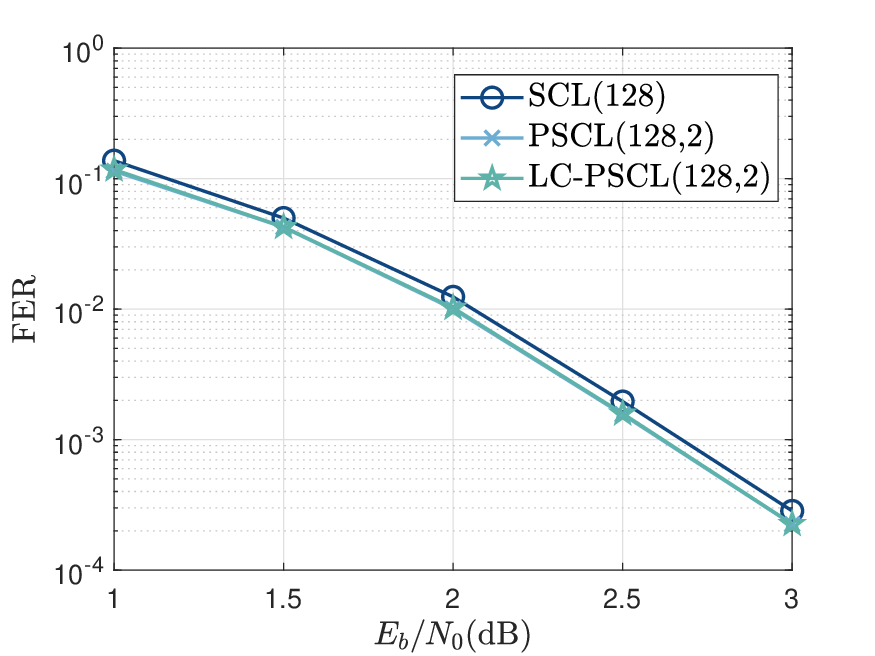}
	\caption{The error performance of different decoding algorithms for a PAC code. Here, $N=128$ and $K=64$.}\label{ebn0FER_N128K64PAC}
	\end{minipage}
	
    \begin{minipage}{\linewidth}
	\centering
	\includegraphics[width=\textwidth]{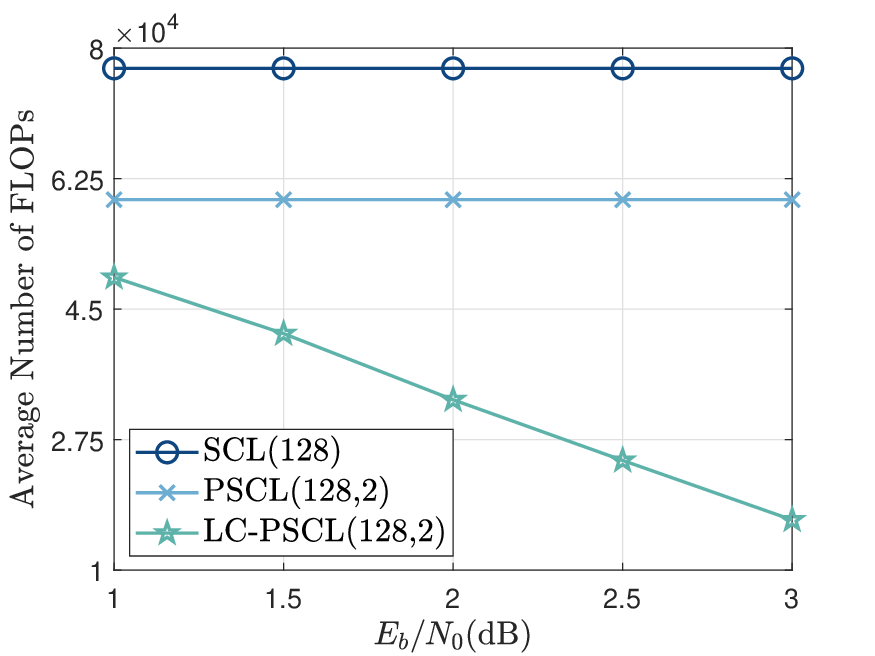}
	\caption{The computational complexity of different decoding algorithms for a PAC code. Here, $N=128$ and $K=64$.}\label{ebn0FLOP_N128K64PAC}
   \end{minipage}	
   	
   \begin{minipage}{\linewidth}
   	\centering
   	\includegraphics[width=\textwidth]{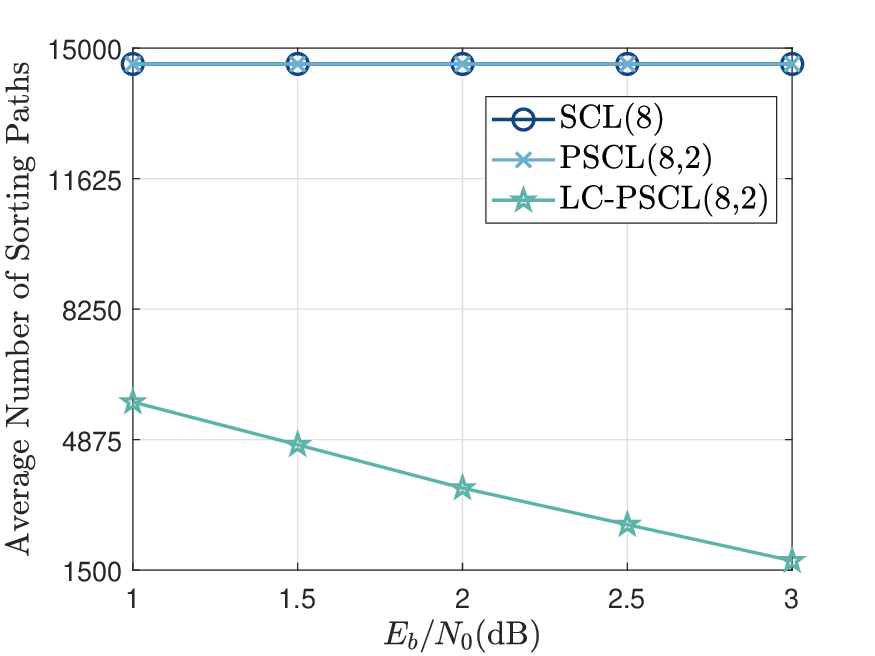}
   	\caption{The sorting complexity of different decoding algorithms for a PAC code. Here, $N=128$ and $K=64$.}\label{ebn0SP_N128K64PAC}
   \end{minipage}
   \vskip0.5cm
\end{figure}

	\begin{example}
	In fact, our proposed LC-PSCL decoder can also be applied to CRC-polar codes and PAC codes. In this example, we simulate a CRC-polar code and a PAC code, each with a code length of $N=128$ and a code dimension of $K=64$. For the CRC-polar code, the information set is determined based on the 5G sequence, and the CRC generator polynomial is $x^{11} + x^{10} + x^{9} + x^{5} + 1$. For the PAC code, the information set is determined based on the Reed-Muller rule, and the impulse response is $(1, 0, 1, 1, 0, 1, 1)$. In addition, for the LC-PSCL, we set $\lambda=0.001$ in both cases. The simulation results for the CRC-polar code are shown in Figs.~\ref{ebn0FER_N128K64CRC11}-\ref{ebn0SP_N128K64CRC11}, and those for the PAC code are provided in Figs.~\ref{ebn0FER_N128K64PAC}-\ref{ebn0SP_N128K64PAC}. These results clearly demonstrate that, compared to the conventional PSCL decoder and SCL decoder, the LC-PSCL decoder achieves reduced computational and sorting complexity with only a minor sacrifice in error performance.
	\end{example}
	
\begin{figure}[!t]
	\begin{minipage}{\linewidth}	\centering
		\includegraphics[width=\textwidth]{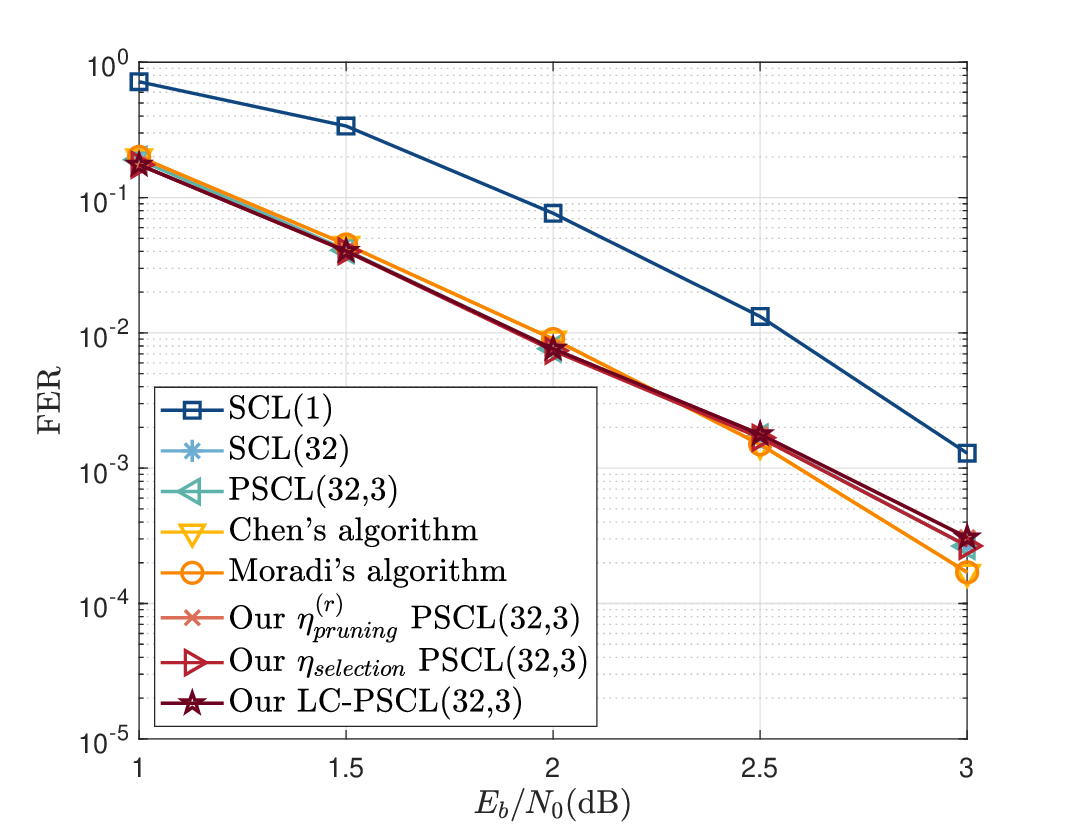}
		\vskip-0.1cm
		\caption{The error performance of different decoding algorithms. Here, $N=1024$ and $K=512$.}\label{ebn0FER_N1024K512L32}
	\end{minipage}
	
	\begin{minipage}{\linewidth}
		\centering	\includegraphics[width=\textwidth]{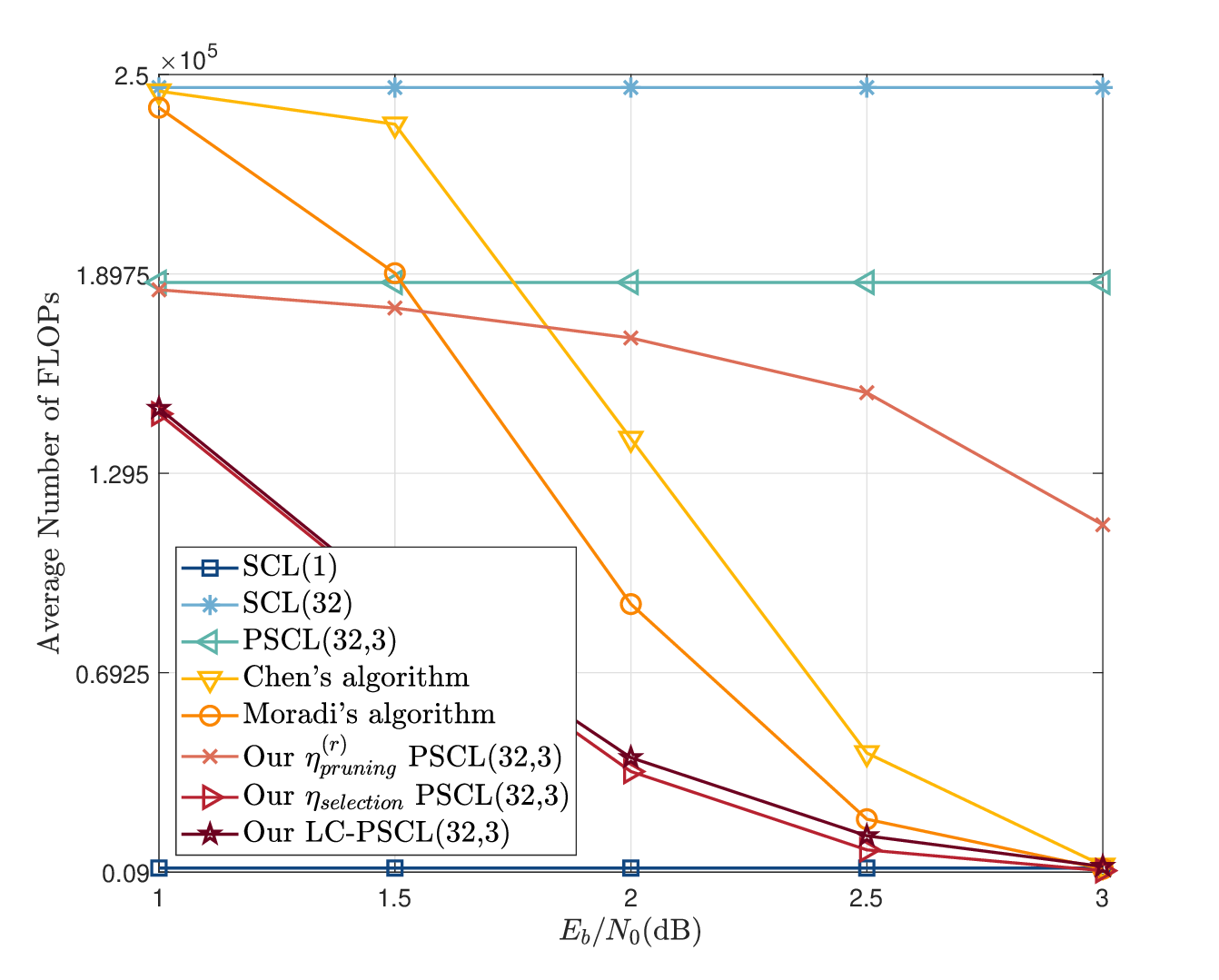}
		\vskip-0.1cm
	    \caption{The computational complexity of different decoding algorithms. Here, $N=1024$ and $K=512$.}\label{ebn0FLOP_N1024K512L32}
	\end{minipage}		
	
	\begin{minipage}{\linewidth}
		\centering	\includegraphics[width=\textwidth]{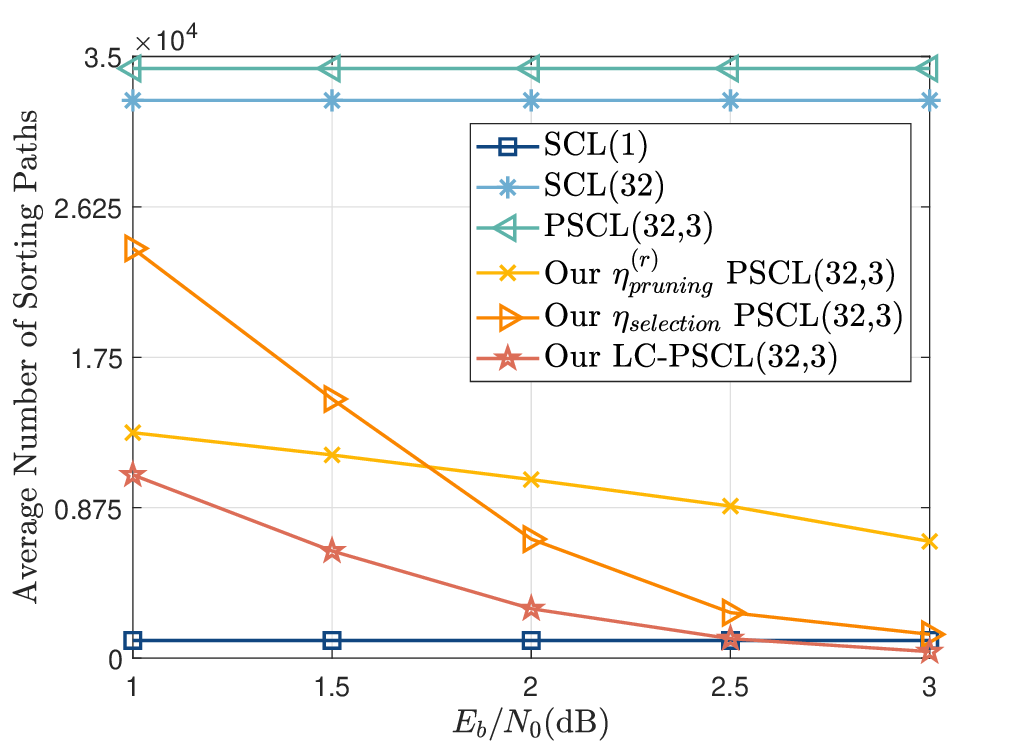}
		\vskip-0.1cm
		\caption{The sorting complexity of different decoding algorithms. Here, $N=1024$ and $K=512$.}\label{ebn0SP_N1024K512L32}
	\end{minipage}			
   \end{figure}		

	\begin{figure*}[!t]
	\centering	\includegraphics[width=\textwidth]{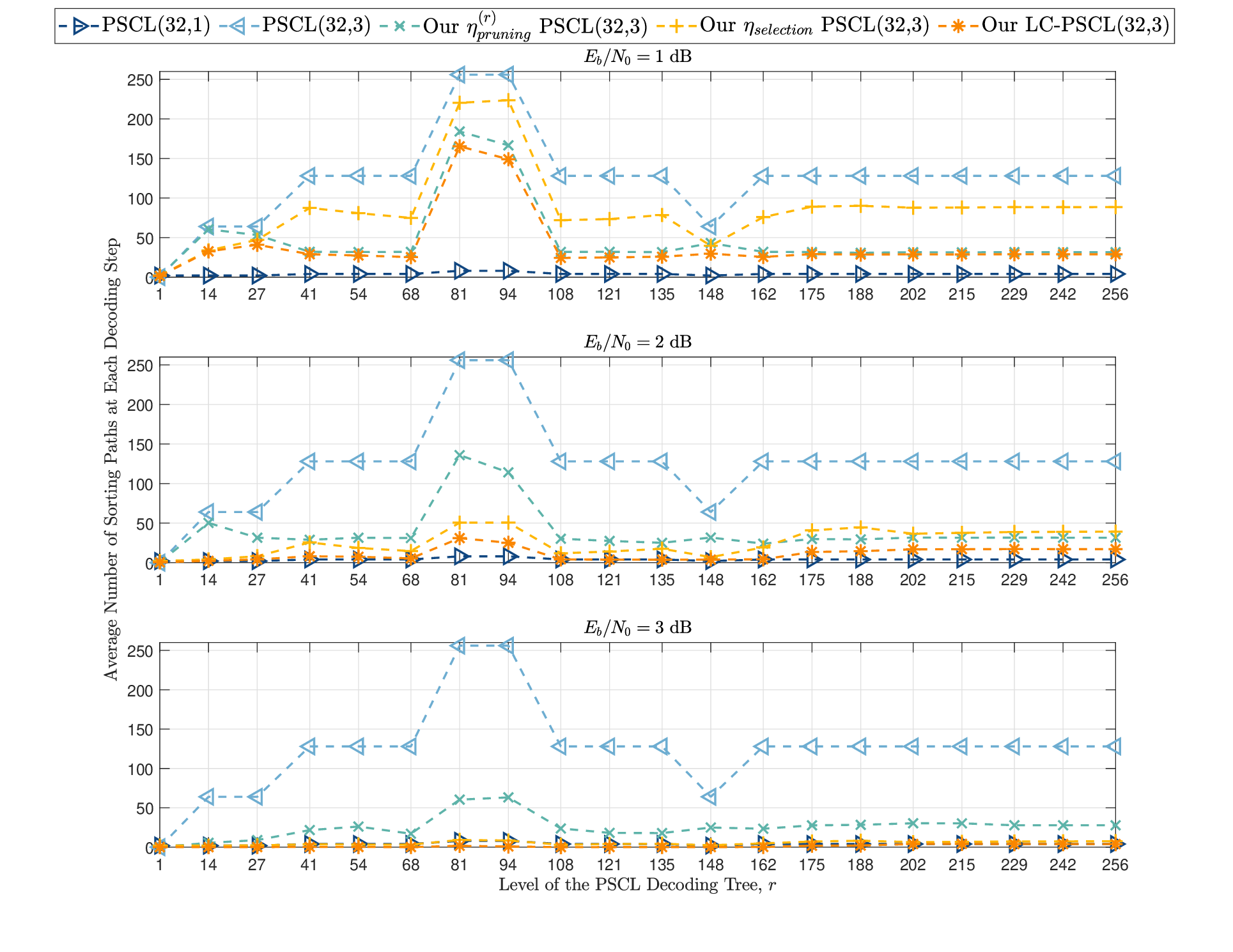}
	\caption{The average number of sorting paths at each decoding step for various decoding algorithms. Here, $N=1024$ and $K=512$, and the PSCL decoding tree associated with the polar code has 256 levels. }\label{ebn0SPatEachDecodingStep_N1024K512L32}	
   \end{figure*}
{
	\begin{example}
	Note that our proposed LC-PSCL decoder incorporates two complexity reduction strategies. One is for the pruning step and another is for the selection step. In this example, we conduct an ablation study to examine the individual effect of each strategy. For this study, we simulate a polar code of length $N=1024$ and dimension $K=512$, which is constructed by the density evolution with Gaussian approximation. The simulation results are presented in Figs.~\ref{ebn0FER_N1024K512L32}-\ref{ebn0SPatEachDecodingStep_N1024K512L32}. In the legend, we denote by ``$\eta_{\text{pruning}}^{(r)}$ PSCL'' the PSCL decoding algorithm modified only at the pruning step and by ``$\eta_{\text{selection}}$ PSCL'' the PSCL decoding algorithm modified only at the selection step. We see that,  compared with the conventional PSCL decoder, both the $\eta_{\text{pruning}}^{(r)}$ PSCL decoder and the $\eta_{\text{selection}}$ PSCL decoder reduce computational/sorting complexity. In terms of the computational complexity, the $\eta_{\text{selection}}$ PSCL decoder is more effective. Regarding the sorting complexity, the $\eta_{\text{pruning}}^{(r)}$ PSCL decoder performs better in the low SNR region, while the $\eta_{\text{selection}}$ PSCL decoder performs better in the high SNR region.
	
	We also compare our proposed two strategies with some existing solutions. The $\eta_{\text{pruning}}^{(r)}$ PSCL decoder exhibits lower computational complexity than Moradi's algorithm~(a pruning method proposed in~\cite{Moradi2023}) in the low SNR region and the $\eta_{\text{selection}}$ PSCL decoder exhibits lower computational complexity than Chen's algorithm~(a selection method proposed in~\cite{Chen2013}). 
	
	Additionally, although the LC-PSCL decoder has a computational complexity similar to the $\eta_{\text{selection}}$ PSCL decoder, it achieves lower sorting complexity by employing our proposed complexity reduction strategy for the pruning step.
	\end{example}
}
	\section{Conclusion}\label{section5}
	In this paper, we introduced a double-threshold strategy for the conventional PSCL decoder. Firstly, we defined a new reliability metric for decoding paths and proposed pruning unreliable paths with reliability metrics less than a preset pruning threshold at each decoding step. Here, the pruning threshold can be efficiently calculated using the GA-based approximation method. Secondly, we assigned a metric to each potential list of paths for estimating the likelihood that the correct path is included in the list, and proposed to select the smallest list among all the lists that meets a preset selection threshold condition. Simulation results show that, the proposed LC-PSCL decoder has the capability to empower polar codes and their generalizations, such as CRC-polar codes and PAC codes, to deliver satisfactory performance with low complexity and latency.
	
	\ifCLASSOPTIONcaptionsoff
	\newpage
	\fi
	\bibliographystyle{IEEEtran}
	\bibliography{IEEEabrv,RefYaoxym}

\end{document}